\documentclass[fleqn,usenatbib,useAMS]{mnras}

\usepackage{graphicx}	
\usepackage{amsmath}	
\usepackage{amssymb}	
\usepackage{multicol}        
\usepackage{bm}		
\usepackage{pdflscape}	

\DeclareMathOperator{\sech}{sech}

\usepackage[T1]{fontenc}
\usepackage{ae,aecompl}

\usepackage{newtxtext,newtxmath}

\title[Evolution of the pseudo bulges]{Growth of disc-like pseudo-bulges in SDSS DR7 since $z=0.1$}

\author[Kumar \& Kataria 2022]{
Ankit Kumar,$^{1,2}$\thanks{E-mail: ankit4physics@gmail.com (AK)}
and Sandeep Kumar Kataria$^{3,4}$\thanks{E-mail: skkatara.iit@gmail.com (SKK)}
\\
$^{1}$Indian Institute of Astrophysics, Bengaluru, 560034, India\\
$^{2}$Joint Astronomy Program, Department of Physics, Indian Institute of Science, Bengaluru, 560012, India \\
$^{3}$School of Physics and Astronomy, Shanghai Jiao Tong University,No.800, Dongchuan Road, Minhang District, Shanghai, 200240, China. \\
$^{4}$Key Laboratory for Particle Astrophysics and Cosmology (MOE) / Shanghai Key Laboratory for Particle Physics and Cosmology, Shanghai 200240, China\\
}

\date{Accepted XXX. Received YYY; in original form ZZZ}

\pubyear{2022}

\begin{document}

\maketitle
\begin{abstract}
Cosmological simulations predict more classical bulges than their observational counterpart in the local Universe. Here, we quantify evolution of the bulges since $z=0.1$ using photometric parameters of nearly 39,000 unbarred disc galaxies from SDSS DR7 which are well represented by two components. We adopted a combination of the S\'ersic index and Kormendy relation to separate classical bulges and disc-like pseudo-bulges. We found that the fraction of pseudo-bulges (classical bulges) smoothly increases (decreases) as the Universe gets older. In the history of the Universe, there comes a point ($z \approx 0.016$) when classical bulges and pseudo-bulges become equal in number. The fraction of pseudo-bulges rises with increasing bulge to disc half-light radius ratio until R$_{\rm e}$/R$_{\rm hlr} \approx 0.6$ suggesting concentrated disc is the most favourable place for pseudo-bulge formation. The mean ellipticity of pseudo-bulges is always greater than that of classical bulges and it decreases with decreasing redshift indicating that the bulges tend to be more axisymmetric with evolution. Also, the massive bulges are progressing towards axisymmetry at steeper rate than the low-mass bulges. There is no tight correlation of bulge S\'ersic index evolution with other photometric properties of the galaxy. Using the sample of multi-component fitting of $S^4G$ data and $N-$body galaxy models, we have verified that our results are consistent or even more pronounced with multi-component fitting and high-resolution photometry.
\end{abstract}


\section{Introduction}
\label{sec:intro}
The galaxy evolution is broadly governed by two kind of processes namely a) gravitational clustering i.e. collapses, merger events and b) internal secular processes like bar, spiral arms etc \citep{Norman.etal.1996, Conselice.2014}. It is well known that internal secular evolution of disc galaxies leads to significant changes in the properties of the bulges \citep{Kormendy.2004, Combes.2009}. Therefore, in order to understand the galaxy formation and evolution processes, the study of bulges is quite insightful. The nature of different type of bulges has been explored in simulations \citep{Athanassoula.2005} as well as in observations \citep{Fisher.Drory.2008, Fisher.Drory.2011, Erwin.et.al.2015}. Morphological studies of galaxies show that the bulge to disc ratio varies from early to late type spiral galaxies in the Hubble sequence \citep{Laurikainen.et.al.2007, Graham.Worley.2008}.

There are broadly two type of bulges given the recent understanding; classical and disc-like bulges. These are differentiated with the help of photometric, kinematic properties and stellar population of the stars they possess \citep{Athanassoula.2005, Fisher.Drory.2008, Athanassoula.2016, Laurikainen.Salo.2016}. Classical bulges are thought to be formed in major mergers \citep{Kauffmann.et.al.1993, Baugh.et.al.1996, Hopkins.et.al.2009, Naab.et.al.2014}, accretion of smaller satellites \citep{Aguerri.2001}, multiple minor mergers \citep{Bournaud.2007, Hopkins.2010}, and monolithic collapse of a primordial cloud \citep{Eggen.1962}. Classical bulges are rounder objects like elliptical galaxies and contains older population star with higher velocity dispersion compare to disc stars \citep{Kormendy.2004}. On the other hand, disc-like bulges are flattened systems like an exponential disc in the nuclear region \cite{Athanassoula.2005}. They are thought to be formed by the inward pulling of gas along the orbits and the consequent star formation \citep{Kormendy.1993, Heller.1994, Regan.2004}. These disc-like bulges are also known as pseudo-bulges \citep{Kormendy.2004} which were conceptualized by \cite{Kormendy.1982,Kormendy.1983}.

Now, the term pseudo-bulge is commonly used to describe disc-like bulges and boxy/peanut structures seen in edge-on galaxies \citep{Athanassoula.2005}. Boxy/peanut structure are vertically thick systems and are dominated by rotational motion of the stars similar to the disc-like bulges. However, it is well-established that Boxy/Peanut structures are just bars seen under different galaxy inclinations \citep{Bureau.Freeman.1999, Lutticke.etal.2000}. They are thought to be formed by disc instability during secular evolution \citep{Kormendy.2004}, vertical heating of the bar due to buckling \citep{Combes.1990, Raha.1991, Martinez.2006, Shen.et.al.2010, Kataria.Das.2018, Kumar.etal.2022}, or heating of the bar due to vertical resonances \citep{Pfenniger.1990}. Classical bulges are very stable against galaxy flybys but boxy/peanut structures grow significantly in major flybys \citep{Kumar.etal.2021}. Both, disc-like pseudo bulges and box/peanut structures, are very different in terms of physical properties and formation mechanism \citep{Athanassoula.2005, Laurikainen.Salo.2016}. To be more specific, we focus our study on classical bulges and disc-like pseudo-bulges (hereafter, we refer disc-like pseudo-bulge as pseudo bulge in through out the draft.)

Magneto-hydrodynamics zoom-in cosmological simulations "Auriga simulations" with sub-grid physics \citep{Gargiulo.et.al.2019} show that pseudo-bulges are prominent in Milky Way type halos. In the same study, it has been shown that around 75 $\%$ of the bulges in these simulations have in-situ stars which are formed around $z=0$ rather than in accretion events. \cite{Fisher.2006} used the PAHs (polycyclic aromatic hydrocarbons) surface brightness profiles and found that the star formation mechanisms for classical bulges and pseudo-bulges are very different. Star formation in classical bulges is fast episodic, whereas pseudo-bulges show long lasting star formation. These studies resonate with semi-analytical modeling of L-galaxies in Millenium and Millenium II simulations \citep{Izquierdo.et.al.2019} which has shown the quiet merger history for pseudo-bulged galaxies compare to classical one.

\cite{Laurikainen.et.al.2007} has shown that pseudo-bulges are widespread along all the Hubble sequence galaxies like classical bulges. It has been also pointed out that pseudo-bulges are found mostly in galaxies with lower bulge to total mass ratio ($B/T$), and galaxies with $B/T$ $\geq$ 0.5 claimed to have classical bulges surely \citep{Kormendy.Fisher.2005}. Several observational studies \citep{Weinzirl.et.al.2009,Kormendy.et.al.2010} find that the fraction of pseudo-bulges in their nearby galaxies sample is larger than 0.5. This fact has raised questions regarding hierarchical structure formation scenario under standard $\Lambda CDM$ cosmology \citep{Kormendy.et.al.2010}.

As we have seen the evolution of pseudo-bulges with cosmic time still remains a mystery. We are motivated to ask questions such as "How do the pseudo-bulge and classical bulge fraction varies with redshift?", "How do the pseudo-bulge properties varies with redshift?". In this article, we study the evolution of bulges where we want to look at the frequencies of the two types of bulges with redshifts. This will surely lead us to understand the connection between two types of bulges.

The plan of the paper is as follows. In Section~\ref{sec:data_analysis}, we have mentioned about the complete sample and our selection criterion along with the $N-$body modeling of galaxies. The evolution of the bulges, their shapes, correlation with various photometric parameters of galaxies, and comparison with local volume survey are shown in Section~\ref{sec:results}. The effect of telescope resolution and disc inclination using simulated galaxies is explored in Section~\ref{sec:effect_of_inc_res}. The effect of data selection criteria on our results is described in Section~\ref{sec:effect_of_data_selection}. In Section~\ref{sec:discussion}, implication of this study are discussed. The brief concluding summary is pointed out in Section~\ref{sec:summary}.

\section{Data and Sampling}
\label{sec:data_analysis}
\subsection{Complete Data}
\label{sec:complete_data}
In this work, we use the archival data from \cite{Simard2011}. The detailed information of data and analysis can be found in \cite{Simard2002, Simard2011} but, for the benefit of the reader, we are describing in brief about the data. \cite{Simard2011} provides the two-dimensional decomposition of 1.12 million objects in g- and r-band from Legacy area of Sloan Digital Sky Survey Data Release Seven (SDSS DR7) \citep{Abazajian2009}. Morphologically, these objects are galaxies and have galactic extinction corrected r-band Petrosian magnitude in the range 14 to 18. The structural parameters of the galaxies were determined using 2D decomposition tool GIM2D \citep{Simard2002}. Only two components, S\'ersic bulge and exponential disc, were used for the decomposition of all the galaxies in the sample. The data is available in three formats of the fittings: (1) fixed S\'ersic index $n=4$ + disc fitting, (2) free S\'ersic index from 0.5 to 8 + disc fitting, and (3) free S\'ersic index from 0.5 to 8 fitting. For the calculation of physical parameters of the galaxies, the cosmological parameters $H_{0}=70$ km s$^{-1}$ Mpc$^{-1}$, $\Omega_{\rm m}=0.3$, $\Omega_{\Lambda}=0.7$ are used.

We also used the archival data from \cite{Salo2015}. The detailed information of data and methodology is described in \cite{Munoz-Mateos.etal.2015, Salo2015}. They provide the two-dimensional decomposition of 2352 galaxies at $3.6~\mu m$ (mid-infrared) form Spitzer Survey of Stellar Structure in Galaxies ($S^4G$) \citep{Sheth2010}. The structural parameters of the galaxies were calculated using 2D decomposition tool GALFIT \citep{Peng2002, Peng2010}. Five components, exponential disc, edge-on disc, S\'ersic bulge, ferrer bar, and unresolved central component psf (point spread function), were used to decompose the galaxies whenever required. The data is available in two formats of the fitting: (1) multi-component fitting, and (2) free S\'ersic index from 0.3 to 19 fitting.

\subsection{Our Criteria}
\label{sec:our_criteria}
\begin{figure*}
    \centering
    \includegraphics[width=\textwidth]{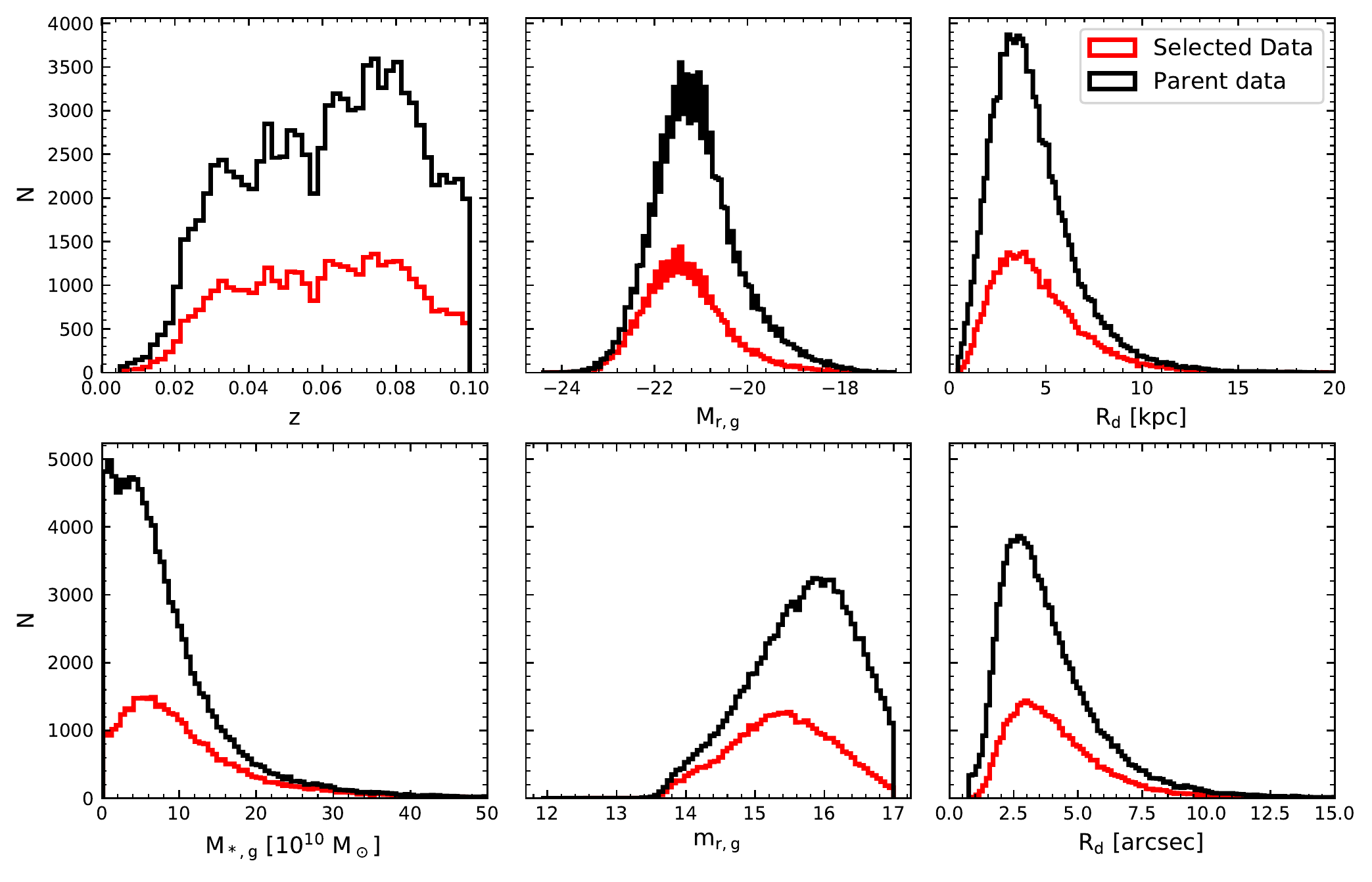}
    \caption{Complete data of galaxies and selected data for this study from \protect\cite{Simard2011}. Data distribution from left to right, top row: spectroscopic redshift ($z$), r-band absolute magnitude of galaxies (M$_{\rm r,g}$), physical disc scale radius (R$_{\rm d}$); bottom row: stellar mass of galaxies (M$_{\rm *,g}$), r-band apparent magnitude of galaxies (m$_{\rm r,g}$), and apparent disc scale radius (R$_{\rm d}$). Black and red histograms show complete data and selected data respectively. Large difference in complete data and selected data is the result of our stringent selection criteria which mostly remove faint and low-mass galaxies.}
    \label{fig:sample_sdss}
\end{figure*}

To understand the evolution of bulges in disc galaxies, we will use the free S\'ersic index from 0.5 to 8 + disc fitting of \cite{Simard2011} (hereafter SDSS data) and multi-component fitting from \cite{Salo2015} (hereafter $S^4G$ data). The redshift or distance of nearby galaxies is severely contaminated by their peculiar velocities. Therefore, we have excluded a small fraction of galaxies with $z<0.005$ as suggested in \cite{Shen2003}. We have limited our study to maximum redshift of $z=0.1$, and spectroscopically confirmed galaxies which are best fitted by two components. We have avoided all the galaxies with apparent bulge effective radius and disc scale radius equal or smaller than the psf radius as these galaxies will have more uncertainties in the photometric parameters. Next we rejected fainter (m$_{\rm r}$ > 17) and smaller galaxies (M$_{*}$ < 10$^{9}$ M$_{\odot}$) from our sample. Stellar mass of the galaxy (M$_{*}$) was estimated using color-stellar mass-to-light ratio relation (CMLR). Stellar mass-to-light ratio ($\gamma_{*}$) of galaxies is related to the color of galaxies by the following expression,
\begin{equation}
    \log \gamma_{*}^{\rm j} = a_{\rm j} + b_{\rm j} \times color,
\end{equation}
where $a_{\rm j}$ and $b_{\rm j}$ are two constants calculated for $\rm j^{th}$ imaging band at given color index (e.g. $g-r$ in our case). To calculate r-band stellar mass-to-light ratio, We have adopted values of $a_{\rm j}$ and $b_{\rm j}$ from \cite{Bell.etal.2003}. Recently, \cite{Du.McGaugh.2020} have re-calibrated these constants for the galaxy mass consistency in various stellar mass-to-light ratio estimators. Both studies give same constants in r-band for $g-r$ color. After all these constraints, we end-up with a total sample of 105,160 galaxies. Hereafter, we refer this sample to parent data.

To minimize the uncertainties in our results, we have imposed the strict constraint of maximum statistical error of $10\%$ on the magnitude of each component, bulge S\'ersic index, disc inclination, bulge to total light ratio, bulge ellipticity, and size of each component. We removed all the galaxies whose S\'ersic index is 0.5 or 8.0 and bulge ellipticity is 0.0 or 0.7 to avoid the fitting bias in the sample due to presence of bar, close merger, point source etc \citep{Simard2011}. Finally, we constrain our sample to the galaxies where half-light radius of the bulge is smaller than the half-light radius of the disc. These constraints reduce our parent data of 105,160 galaxies to 40,504 galaxies.

Since we are interested in the evolution of the bulges, we need to take care of barred galaxies in the sample. To remove the barred galaxies from our sample, we used the catalog of "Galaxy Zoo 2" classification \citep{Willett.etal.2013, Hart.etal.2016}. Galaxy Zoo is a citizen science project, where images of the galaxies are shown to the interested citizens and they are asked to answer some basic questions related to the galaxy morphology. These answers are used for morphological classification of the galaxies depending on the probability of the voting. Though this method does not removes all the barred galaxies, but it improves our sample. Removal of barred galaxies reduces our sample from 40,504 to 38,996 galaxies. Hereafter, we refer this sample to selected data.

For a comparison between our selected data and the parent data of galaxies, we have shown the number distribution of galaxies in Fig.~\ref{fig:sample_sdss}. From left to right, top row panels: spectroscopic redshift ($z$), r-band absolute magnitude of galaxies (M$_{\rm r,\rm g}$), physical disc scale radius (R$_{\rm d}$); bottom row panels: stellar mass of galaxies (M$_{*,\rm g}$), r-band apparent magnitude of galaxies (m$_{\rm r,\rm g}$), and apparent disc scale radius (R$_{\rm d}$). Physical size of galaxy was calculated using its angular distance as mentioned in \cite{Simard2011}. Our selected data represents the fair sample of the parent data except at faint and low-mass end of the distribution. It is because our strict constraint of maximum $10\%$ statistical error in each photometric parameter. Later, in Section~\ref{sec:effect_of_data_selection}, we will discuss the effect of this strict selection criteria on our results.

In our analysis, we have divided data in several different categories. The disc dominating and bulge dominating galaxies are defined on the basis of bulge to total light ratio $B/T$ in r-band. The disc dominating galaxies are those which have $B/T \leq 0.5$ and the bulge dominating galaxies have $B/T > 0.5$. None of the photometric classification of the bulges provides proper separation between classical bulges and pseudo-bulges. Particularly, $n=2$ S\'ersic index does not show bi-modality in the distribution of the bulges \citep{Graham.2013, Graham.2014, Costantin.etal.2018, Mendez-Abreu.etal.2018, Gao.etal.2020, Kumar.etal.2021}. Some elliptical galaxies, and bulges of S0 and merger built galaxies also show lower S\'ersic indices \citep{Davies.etal.1988, Young.Currie.1998, Eliche-Moral.2011, Querejeta.etal.2015, Tabor.etal.2017}. In Section~\ref{sec:effect_of_inc_res}, We have also verified using numerical models of Milky Way mass galaxies that S\'ersic index is not reliable for bulge classification, given the inclination of the disc and resolution of the telescope. Hence, for a more precise classification of the bulges, we have adopted a combination of S\'ersic index based classification and Kormendy relation based classification, which we call S\'ersic-Kormendy classification. S\'ersic index based classification (hereafter S\'ersic classification) makes use of $n=2$ S\'ersic index. All the bulges having $n \leq 2$ are categorized as pseudo-bulges, whereas bulges with $n > 2$ are categorized as classical bulges \citep{Fisher2006_proce, Fisher.Drory.2008}. On the other hand, Kormendy relation based classification (hereafter Kormendy classification) uses well known Kormendy relation for elliptical galaxies to separate two types of bulges. All the bulges which lie above $3\sigma$ limit of the Kormendy relation are grouped as classical bulges, while the bulges which fall below $3\sigma$ limit are grouped as pseudo-bulges \citep{Gadotti.2009}. For this study, we have taken Kormendy relation from \cite{Lackner.Gunn.2012}. In S\'ersic-Kormendy classification, only those bulges are considered which satisfy both S\'ersic and Kormendy classifications simultaneously. For the sake of comparison, we will also be showing our results using S\'ersic classification and Kormendy classification along with S\'ersic-Kormendy classification.

\subsection{Simulated Galaxy Models}
\label{sec:galaxy_models}
To understand the effect of spatial resolution of telescope, projection of galaxy, and surface density of galaxy on fitting parameters, particularly S\'ersic index, we simulate model galaxies with bulge to disc mass ratio ($B/T$) 0.1, 0.3, 0.5, and 0.7 keeping fixed disc mass and fixed bulge scale radius. There are four disc surface densities for each model. The total mass (stellar bulge + stellar disc + dark halo) of each the model galaxy is similar to that of Milky Way type galaxy. We use the publicly available $N-$body code GALIC \citep{Yurin2014} to generate our model galaxies. In our models, dark matter halo density is represented by Hernquist profile,
\begin{equation}
    \rho_{\rm dm}(r)=\frac{M_{\rm dm}}{2\pi}\frac{a}{r(r+a)^3}
    \label{eqn:halo}
\end{equation}
where `$a$' is dark matter halo scale radius. This is related to the concentration parameter `$c$' of a corresponding NFW halo \citep{NFW1996} of mass M$_{\rm dm}$=M$_{\rm 200}$ by the following expression,
\begin{equation}
    a=\frac{r_{200}}{c}\sqrt{2\left[\ln{(1+c)-\frac{c}{(1+c)}}\right]}
    \label{eqn:scale_radius}
\end{equation}
where r$_{200}$ is the virial radius of galaxy. This is the radius within which the average matter density is 200 times the critical density of the Universe. M$_{200}$ is mass within the virial radius.

The disc density decays exponentially in the radial direction and its vertical distribution is described using $\sech^{2}$ profile
\begin{equation}
    \rho_{\rm d}(\rm R,\rm z)=\frac{M_{\rm d}}{4\pi z_{0} R_{\rm s}^{2}}\exp\left(-{\frac{R}{R_{\rm s}}}\right) \sech^{2}\left(\frac{z}{z_{0}}\right)
    \label{eqn:disc}
\end{equation}
where M$_{\rm d}$ is total disc mass, R$_{\rm s}$ is disc scale radius, and $z_{0}$ is disc scale height.

The bulge density is also modelled by the spherically symmetric Hernquist density profile
\begin{equation}
    \rho_{\rm b}(r)=\frac{M_{\rm b}}{2\pi}\frac{b}{r(r+b)^{3}}
    \label{eqn:bulge}
\end{equation}
where M$_{\rm b}$ represents the total bulge mass and `$b$' represents the bulge scale radius.

We set 1 million particles for each component which results in a total of 3 million particles in each model galaxy. The rotation velocity of each model galaxy is set to be 220 km s$^{-1}$ which corresponds to a total mass $2.47\times10^{12}$M$_{\odot}$ (where M$_{\odot}$ = mass of the Sun). The disc mass is fixed at 0.04 of the total galaxy mass. The gravitational softening is set to be 0.01 $kpc$ for each type of particle.

For the purpose of bulge-disc decomposition of the simulated galaxies, we use the latest version of GALFIT \citep{Peng2002, Peng2010} which is a widely used for 2d decomposition of galaxies. To see the effect of the spatial resolution, the pixel size is set to be $0.025~kpc\times0.025~kpc$ (hereafter high-resolution) and $0.05~kpc\times0.05~kpc$ (hereafter low-resolution). For the decomposition of simulated galaxies, we did not consider any background sky and the point spread function (psf) is taken to be delta function.

\section{Results}
\label{sec:results}
\subsection{Distribution of Bulges with Redshift}
\label{sec:bulge_with_z}
\begin{figure*}
    \centering
    \includegraphics[width=\textwidth]{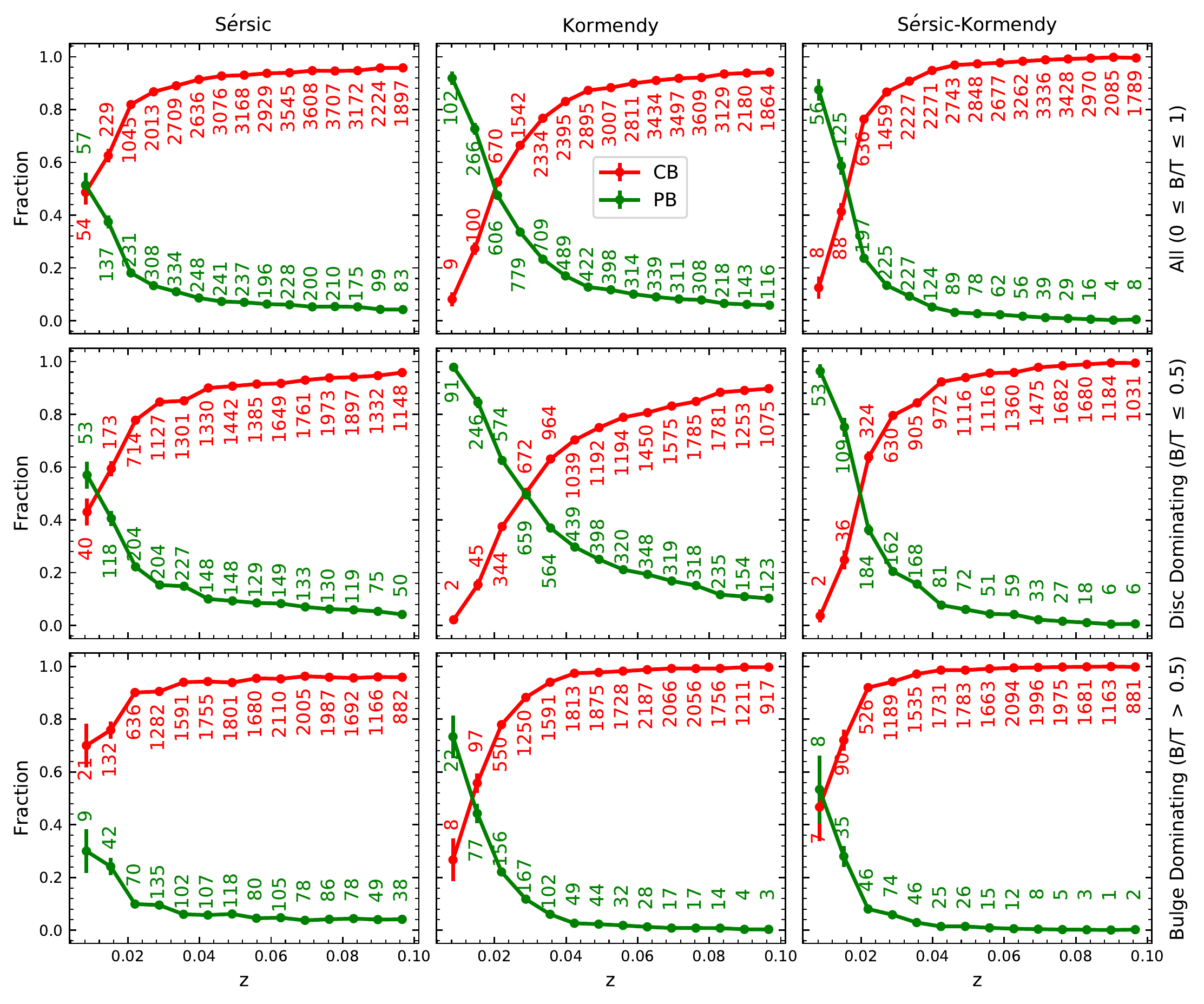}
    \caption{Fractional distribution of the classical and pseudo-bulges with redshift ($z$) in the SDSS survey. Left, middle, and right columns show S\'ersic, Kormendy, S\'ersic-Kormendy classifications respectively. Top, middle, and bottom rows represent all, disc dominating, and bulge dominating galaxies respectively. Red and green curves represent classical bulges (CB) and pseudo-bulges (PB) respectively. The numbers of classical bulge and pseudo-bulge galaxies at any point are shown with their respective colors. All the methods show domination of pseudo-bulges over classical bulges with decreasing redshift.}
    \label{fig:bulge_with_z}
\end{figure*}

The evolution of the classical and pseudo-bulges over the cosmic time from redshift 0.005 to 0.1 is represented in the Fig.~\ref{fig:bulge_with_z} from SDSS survey. The effect of classification method on the distribution of the bulge type is shown in the left, middle, and right columns of the figure which are corresponding to the S\'ersic classification, Kormendy classification, and S\'ersic-Kormendy classification respectively. The contribution of galaxy type on the distribution of the bulge type is shown in the top, middle, and bottom rows of the figure which are the representative of all, disc dominating, and bulge dominating galaxies respectively. Red curves show fraction of classical bulges and green curves show fraction of pseudo-bulges. Statistical uncertainty in the fraction is shown using error bars and is estimated from $\sqrt{[f \times (1-f)]/N}$, where $f$ is the fraction of any point and $N$ is number of objects at that point in the distribution \citep{Sheth.etal.2008}. The numbers of classical bulge and pseudo-bulge galaxies at any point are shown with their respective colors. At many points, error bar is smaller than the size of the symbol. All the panels (except bottom left) of this figure indicate that the local Universe is dominated by pseudo-bulges. The high fraction of the pseudo-bulges is contrary to the prediction of cosmological simulations. According to widely accepted $\Lambda CDM$ cosmological model, we expect more classical bulges instead of pseudo-bulges \citep{White.Rees.1978, Aguerri.2001, Bournaud2005, Baugh.2006, Brooks2016}.

As we go back in the time towards the high-redshift Universe, the fraction of pseudo-bulge decreases and classical bulges start dominating over the pseudo-bulges. In the evolution of the Universe, there comes a time before that Universe was dominated by the classical bulges or spheroids. Since no bulge classification method shows perfect bimodality in two types of bulges, the precise redshift of equality in classical bulges and pseudo-bulges cannot be well constrained. S\'ersic classification shows the point of equality a lower redshift than the Kormendy classification. As a consequence, S\'ersic-Kormendy classification show equality in between the two classification. If we see the distribution of all the sample galaxies in S\'ersic, Kormendy, and S\'ersic-Kormendy classifications, the fraction of the classical and pseudo-bulge becomes equal at $z \approx$ 0.009, 0.020, 0.016 redshifts respectively.

One can see the effect of bulge classification criterion on the distribution of bulge type in all, disc dominating, and bulge dominating galaxies by comparing the panels of top, middle, and bottom row respectively. All three rows show that the fractional distribution of the bulges in S\'ersic classification is very different than the other two classifications. In the low-redshift region, it always shows small fraction of pseudo-bulges as compare to the Kormendy and S\'ersic-Kormendy classifications. But, the trends of distribution are more or less similar in Kormendy and S\'ersic-Kormendy classifications except for bulge dominating galaxies at low-redshift where Kormedy classification show more pseudo-bulges than the S\'ersic-Kormendy classification.

Similarly, the contribution of galaxy type on the distribution of bulge type in S\'ersic, Kormendy, S\'ersic-Kormendy classifications can be seen by comparing the panels of left, middle, and right column respectively. From all three columns, it is clear that the disc dominating galaxies provide more pseudo-bulges than the bulge dominating galaxies in the fractional distribution of the bulges at low-redshift. This implies that most of the pseudo-bulges are low mass relative to their hosting discs. Hence, we conclude that the Kormendy and S\'ersic-Kormendy classifications show quite similar distribution of the bulges but, S\'ersic classification results in a lower fraction of pseudo-bulges at low-redshift. Kormendy classification show slightly higher fraction of pseudo-bulges than S\'ersic-Kormendy classification at low-redshift in bulge domination galaxies. However, local volume remains pseudo-bulge domination irrespective to the classification scheme, and the dominating contribution of pseudo-bulges comes, mostly, from the disc dominating galaxies.

\subsection[]{Distribution of Bulges with R$_{\rm e}$/R$_{\rm hlr}$}
\label{sec:bulge_with_ReRd}
\begin{figure*}
    \centering
    \includegraphics[width=\textwidth]{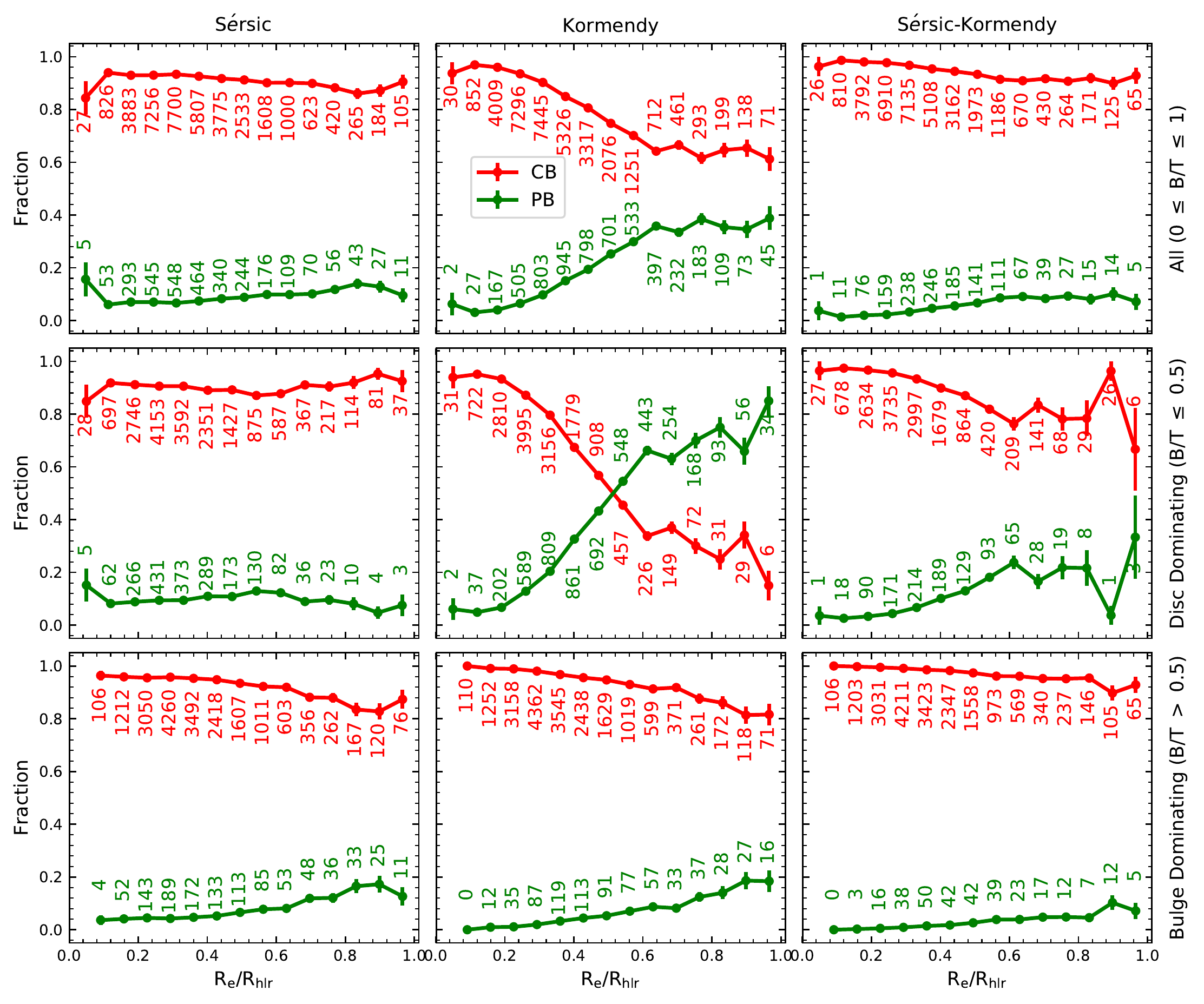}
    \caption{Fractional distribution of classical and pseudo-bulges with the ratio of bulge semi-major half-light radius (R$_{\rm e}$) to disc semi-major half-light radius (R$_{\rm hlr}$) in the SDSS survey. Left, middle, and right columns show S\'ersic, Kormendy and S\'ersic-Kormendy classifications respectively. Top, middle, and bottom rows represent all, disc dominating, and bulge dominating galaxies respectively. Red and green curves represent classical bulges and pseudo-bulges respectively. The numbers of classical bulge and pseudo-bulge galaxies at any point are shown with their respective colors. The fraction of pseudo-bulges increase with increasing bulge to disc size ratio.}
    \label{fig:bulge_with_ReRd}
\end{figure*}

To see the effect of the bulge and disc sizes on the distribution of bulge type, we have calculated the fractional distribution of the classical and pseudo-bulges with the ratio of bulge semi-major half-light radius R$_{\rm e}$ to disc semi-major half-light radius (R$_{\rm hlr}$) from SDSS survey in Fig.~\ref{fig:bulge_with_ReRd}. The ratio of bulge semi-major half-light radius to disc semi-major half-light radius (R$_{\rm e}$/R$_{\rm hlr}$) is better than the absolute bulge or disc size because it removes the error, if any, in the size calculation using distance or redshift of the galaxy. However, in appendix~\ref{app:z_vs_Re_Rd}, we have also shown the absolute sizes of the bulges and discs with redshift and compared them together. In this figure, columns show the effect of the classification method and rows show the effect of galaxy type on the distribution of bulge type. Left, middle, and right columns represent S\'ersic, Kormendy, and S\'ersic-Kormendy classifications respectively. Top, middle, and bottom rows are the representatives of all, disc dominating, and bulge dominating galaxies respectively. From all the panels of the figure, it is clear that a large fraction of the small bulges (small R$_{\rm e}$/R$_{\rm hlr}$) is classical in nature. The fraction of pseudo-bulges increases with increasing R$_{\rm e}$/R$_{\rm hlr}$ or vice-versa the fraction of classical bulges increases with decreasing R$_{\rm e}$/R$_{\rm hlr}$. Unitl R$_{\rm e}$/R$_{\rm hlr} \approx 0.6$, these trends are valid in all the panels.

The effect of bulge classification method on the distribution of bulge type in all, disc dominating, and bulge dominating galaxies can be seen by comparing the panels of the top, middle, and bottom rows respectively. All the three rows of the figure show that the Kormendy classification results in more pseudo-bulges than the other two classification methods for R$_{\rm e}$/R$_{\rm hlr} \geq 0.6$. Particularly, this difference is more distinct in disc dominating galaxies. However, the classification of bulges in bulge dominating galaxies does not shows significant change among three methods. In the same way, one can see the effect of galaxy type on the distribution of bulge type in S\'ersic, Kormendy, and S\'ersic-Kormendy classifications by comparing the panels of left, middle, and right columns respectively. All the columns of the figure clearly indicate that the contribution of disc dominating galaxies in the distribution of pseudo-bulges is always great or equal to that of the bulge dominating galaxies until R$_{\rm e}$/R$_{\rm hlr} \approx 0.6$. Hence, for R$_{\rm e}$/R$_{\rm hlr} \leq 0.6$, a large fraction of pseudo-bulges comes from the disc dominating galaxies. Inversely, a large fraction of classical bulges comes from the bulge dominating galaxies. These conclusions hold for all the classification methods discussed here. When moving towards larger value of R$_{\rm e}$/R$_{\rm hlr}$, disc dominating galaxies start showing increasing classical bulges in S\'ersic and S\'ersic-Kormendy classifications. But, the Kormendy classification still shows increasing pseudo-bulges in disc dominating galaxies. According to Hubble classification, S0 galaxies have massive and large bulges which are comparable to their discs. These bulges are generally classical in nature. Therefore, we expect increase in the fraction of classical bulges close to R$_{\rm e}$/R$_{\rm hlr} = 1$ which is not coming out in Kormendy classification. However, S\'erisc-Kormendy classification captures it well.

For a given bulge to total light ratio ($B/T$), the distribution of the bulges with R$_{\rm e}$/R$_{\rm hlr}$ follows more or less similar trends to that of Fig.~\ref{fig:bulge_with_ReRd}. This implies that the ratio R$_{\rm e}$/R$_{\rm hlr}$ can be treated as an indicator of the ratio of disc to bulge mean surface densities (see the appendix~\ref{app:bulge_disc_bright}). For example, we can say that the bulges with R$_{\rm e}$/R$_{\rm hlr}<0.5$ are more concentrated (or dense) than the bulges with R$_{\rm e}$/R$_{\rm hlr}>0.5$ relative to their hosting discs. From all the panels of the Fig.~\ref{fig:bulge_with_ReRd}, we can see that the preferential condition for the bulges to be pseudo is their low concentration relative to the hosting discs. Concentrated bulges are usually classical in nature. Hence the relative surface densities of the bulge and disc play a crucial role in the formation and evolution of the bulges. A low surface density disc usually lacks from global instabilities in its center \citep{Toomre.1964, Mihos.etal.1997, Mayer.Wadsley.2004, Sodi.etal.2017, Peters.Rachel.2019}. So, most probably, it will also lack from the pseudo-bulge.

\subsection{Distribution of Bulge Ellipticity with Redshift}
\label{sec:ellipticity_with_z}
\begin{figure*}
    \centering
    \includegraphics[width=\textwidth]{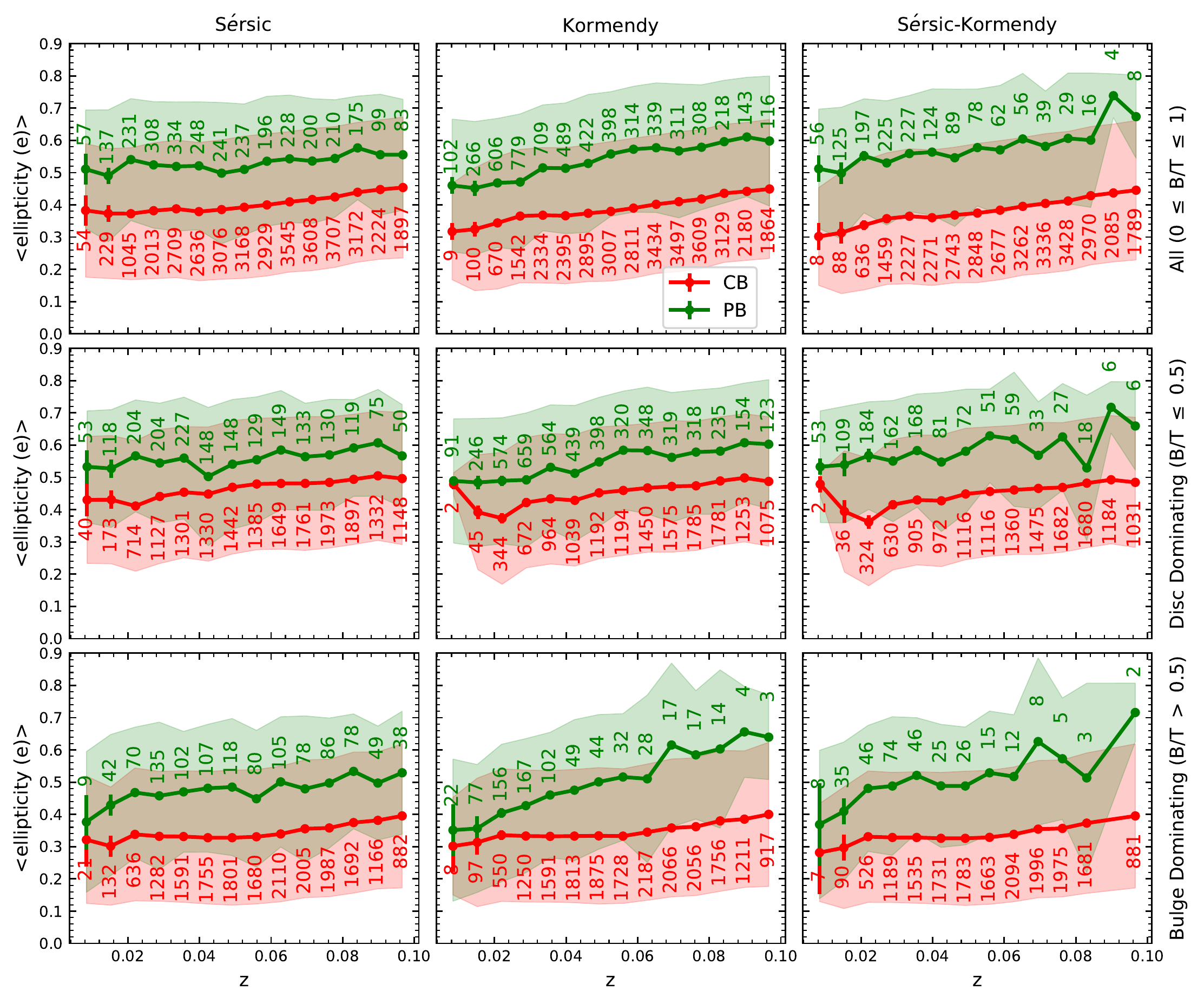}
    \caption{Distribution of mean deprojected bulge ellipticity ($<e=1-b/a>$) with redshift in the SDSS survey. Left, middle, and right columns show S\'ersic, Kormendy, and S\'ersic-Kormendy classifications of bulges respectively. Top, middle, and bottom rows represent all, disc dominating, and bulge dominating galaxies respectively. Red, and green color curves represent the distribution of classical bulges, and pseudo-bulges bulges respectively. Shaded regions show standard deviation from mean and vertical bars represent statistical uncertainty due to number of galaxies. The numbers of classical bulge and pseudo-bulge galaxies at any point are shown with their respective colors. All panels show that bulges are moving close to the round shape with decreasing redshift.}
    \label{fig:ellipticity_with_z}
\end{figure*}

Classical bugles are supported by the random motion of stars that makes them more round in shape as compare to the pseudo-bulges which are supported by the ordered rotational motion of the stars. Rotational motion of the stars in pseudo-bulges provide them flat shape. In its evolution, galaxy goes through several dynamical and morphological changes due to secular evolution and interactions with other galaxies \citep{Kumar.etal.2021}. To trace the morphological changes in the bulges during the evolution of the galaxies, we calculated mean ellipticities ($<e>$) of the classical bulges and pseudo-bulges as a function of redshift. For this purpose, we deprojected all the bulges to minimize the effect of disc inclination on projected shape of the bulges (mainly pseudo-bulges). Fig.~\ref{fig:ellipticity_with_z} shows the evolution of the mean bulge ellipticity of classical bulges and pseudo-bulges with redshift for SDSS data. Red and green curves represent classical bulges and pseudo-bulges respectively. The standard dispersion from mean is displayed using shaded regions around mean values, and statistical uncertainty due the number of galaxies is shown with vertical bars. The numbers of classical bulge and pseudo-bulge galaxies at any point are shown with their respective colors. We have demonstrated S\'ersic, Kormendy, and S\'ersic-Kormendy bulge classification schemes in first, middle, and right columns respectively. Rows, starting from the top, show the ellipticity evolution in all, disc dominating, bulge dominating galaxies respectively.

One thing we can clearly notice from this figure is that the mean ellipticity of the pseudo-bulges is always higher than the classical bulges irrespective to the classification scheme and redshit. It means that the classical bulges are generally rounder than the pseudo-bulges when seen in face-on projection of galaxies. On the other hand, all the panels show decreasing mean ellipticity with decreasing redshift for both types of bulges (when considering points with statistically significant count of galaxies for reasonable mean ellipticity). The reducing mean ellipticity of the bulges implies that the both types of bulges are getting rounder and rounder with the evolution of the galaxies.

The effect of bulge classification scheme on the evolution of mean bulge ellipticity in all, disc dominating, and bulge dominating galaxies can be marked by comparing all the panels of top, middle, and bottom rows respectively. From all the rows, we can see that Kormendy and S\'ersic-Kormendy classification show quite similar declining trends in mean ellipticity for both types of bulges. S\'ersic classification method exhibits shallower decline in mean ellipticity of classical bulges than the other two classification methods. But, at low-redshift, the ellipticity of pseudo-bulges remains nearly unaffected in three classification methods. In the similar manner, one can observe the effect of galaxy type on the distribution of bulge ellipticity in S\'ersic, Kormendy, and S\'ersic-Kormendy classifications by comparing all the panels of left, middle, and right columns respectively. From all the columns, one can clearly see that both types of bulges in bulge domination galaxies show lower mean ellipticity than the disc dominating galaxies. It is true for whole redshift range and for three classifications when considering points with statistically significant number of galaxies. Also, the rate of declining mean ellipticity is steeper in bulge dominating galaxies than in disc dominating galaxies. Now, we can conclude that the shape of massive bulges is more axisymmetric than the low-mass bulges at whole redshift range. Also, the high-mass pseudo-bulges are moving rapidly towards the axisymmetry.

One should note that we have removed the sample of barred galaxies in our analysis. The observed fraction of the barred galaxies ranges from $30\%$ to $60\%$ in the local volume \citep{Aguerri.etal.2009, Masters.etal.2011, Sodi.etal.2015, Diaz-Garcia.etal.2016}. This fraction declines when we move from low-redshift to high-redshift \citep{Sheth.etal.2008, Melvin.etal.2014}. Therefore, it may possible that some galaxies with bulges had formed bar during their evolution. The conversion of bulged galaxies into barred galaxies can affect the mean ellipticity of the bulges. But, how much it will influence evolution of the bulges remains the question for further study.

\subsection{Evolution of the Bulges with Galaxies}
\label{sec:gal_prop_with_z}
\begin{figure*}
    \centering
    \includegraphics[width=\textwidth]{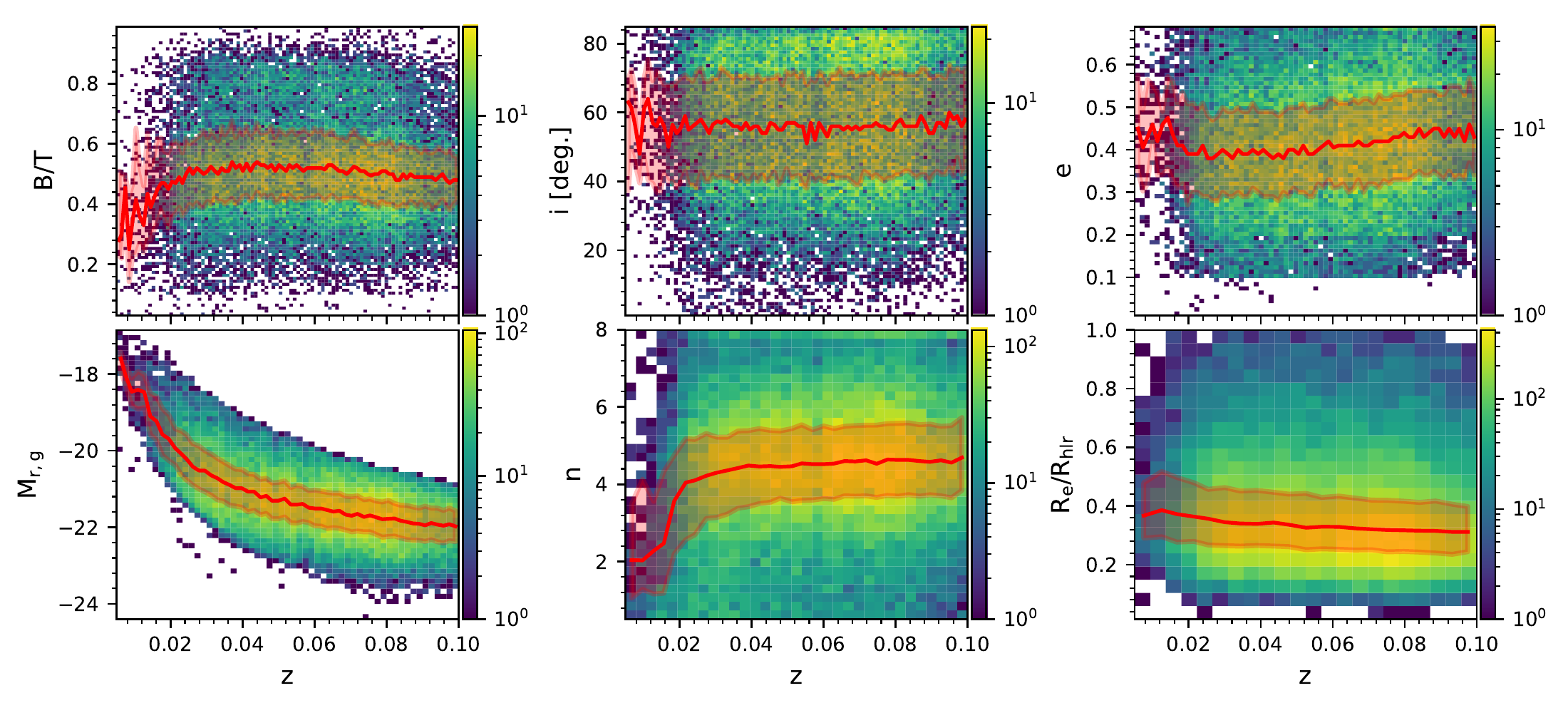}
    \caption{Distribution and evolution of different photometric properties of the galaxies with the redshift ($z$) from SDSS survey. On the y axis, $B/T$ = bulge to total light ratio, $i$ = disc inclination, e = bulge ellipticity, $M_{r,g}$ = r band galaxy absolute magnitude, n = bulge S\'ersic index, and $R_{\rm e}/R_{\rm hlr}$ = bulge to disc half-light radius ratio. The number density of the galaxies is represented by the color on log scale. The red solid curves show the median of the data and shaded region represents the $\pm 25 \%$ dispersion of the data from median.}
    \label{fig:gal_prop_with_z}
\end{figure*}

The different components of the galaxies evolve together with the evolution of the Universe. Therefore, it will be interesting to understand the evolution of the photometric properties of the galaxies with the redshift and their correlations with the evolution of the bulges. In the Fig.~\ref{fig:gal_prop_with_z}, we have shown the distribution and evolution of bulge to total light ratio ($B/T$), disc inclination ($i$), bulge ellipticity ($e$), r band galaxy absolute magnitude (M$_{\rm r,g}$), bulge S\'ersic index ($n$), and bulge to disc half light radius ratio (R$_{\rm e}$/R$_{\rm hlr}$) with redshift ($z$) from SDSS survey. The color of the distribution represents the number density of galaxies on log scale. The red solid curves represents the median of the data and shaded color show the $\pm 25 \%$ dispersion of the data from median. 

From the first column of the first row, one can see that the median of bulge to total light ratio ($B/T$) is decreasing with decreasing redshift. The Universe is becoming disc dominating (dynamically cool) as it is getting older and older. In the local Universe, median + 25$\%$ curve is below the $B/T=0.5$ cut-off i.e. matter fraction in disc component is higher than the bulge component. This suggests that nearly 75$\%$ of the local Universe is dynamically cool. Similar signatures of dynamical cooling can be found from the second and third columns of the first row which show that the medians of inclination ($i$) is nearly constant however the bulge ellipticity ($e$) is increasing with decreasing redshift. Increase in the bulge ellipticity at the constant disc inclination is the indication of the deviation of the bulges from their classical nature. Note that the less number of data points near zero ellipticity in third column of first row is the result of our tight criteria of maximum $10\%$ error in any quantity. Also, the increase in median ellipticity here should not be confused with decrease in mean ellipticity in Fig.~\ref{fig:ellipticity_with_z}. Here, we have considered inclined galaxies.

First column of the second row shows that the median of r band galaxy absolute magnitude (M$_{\rm r,g}$) is increasing with decreasing redshift. These two quantity, magnitude and redshift, have strong and obvious correlation because at a given redshift/distance, we cannot observe an object fainter than the limit of the telescope. Therefore, we are seeing increased number of the fainter galaxies with decreasing redshift. 
The second column of the second row illustrates the decreasing median of S\'ersic index ($n$) with decreasing redshift. This also indicates the dynamical cooling of the galaxies with the evolution of the Universe. The third column of the second row shows the evolution of the bulge to disc semi-major half light radius ratio (R$_{\rm e}$/R$_{\rm hlr}$) with redshift. Though the change is very small, but it is increasing linearly indicating either faster growth of the bulges than the discs or the growing ellipticity of the bulges. 

\begin{figure}
    \centering
    \includegraphics[width=\columnwidth]{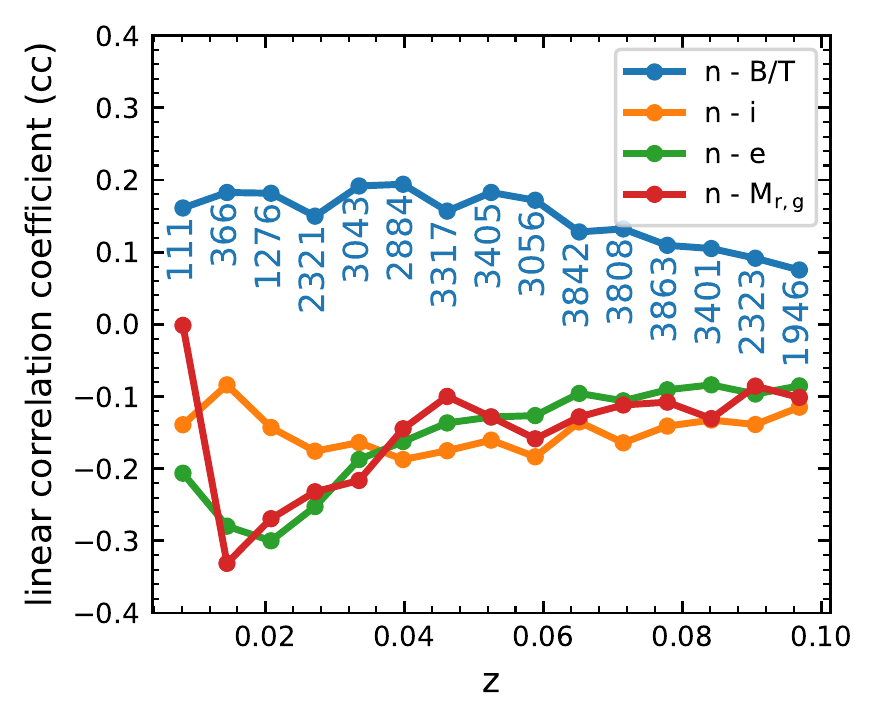}
    \caption{Linear correlation coefficient of the bulge S\'ersic index ($n$) with different photometric properties of the galaxies in different redshift bins of SDSS survey. The blue, orange, green, and red color solid curves represent the linear correlation of bulge S\'ersic index with bulge to total light ratio ($B/T$), disc inclination ($i$), bulge ellipticity ($e$), and r band galaxy absolute magnitude (M$_{\rm r,g}$) respectively. The number of galaxies at any point are also shown at respective redshift.}
    \label{fig:bulge_corr_coef}
\end{figure}

We have also calculated the Pearson's linear correlation coefficient (cc) of the bulge S\'ersic index ($n$) with other photometric properties of the galaxies in different redshift bins of the SDSS survey. The linear correlation coefficient of two variables tells how the one variable changes with the change in the other variable. Its value lies between -1 (for a perfect negative correlation) and +1 (for a perfect positive correlation). In the Fig.~\ref{fig:bulge_corr_coef}, the blue, orange, green, and red color solid curves represent the linear correlation of bulge S\'ersic index with bulge to total light ratio ($B/T$), disc inclination ($i$), bulge ellipticity ($e$), and r band galaxy absolute magnitude (M$_{\rm r,g}$) respectively. At high-redshift, $B/T$ has very weak positive correlation. It weakly develops positive correlation with decreasing redshift. Disc inclination has very weak and nearly constant anti-correlation at all redshifts. Bulge ellipticity and absolute galaxy magnitude show the weakly increasing negative correlation as we move towards the low-redshift Universe. However, they show sudden decrease in the negative correlation at very low-redshift. Only bulge ellipticity and absolute galaxy magnitude show significant, but small, anti-correlation just before the sudden drop at low-redshift.

\subsection{Comparison with Local Volume Survey}
\label{sec:compare_sdss_s4g}
\begin{figure}
    \centering
    \includegraphics[width=\columnwidth]{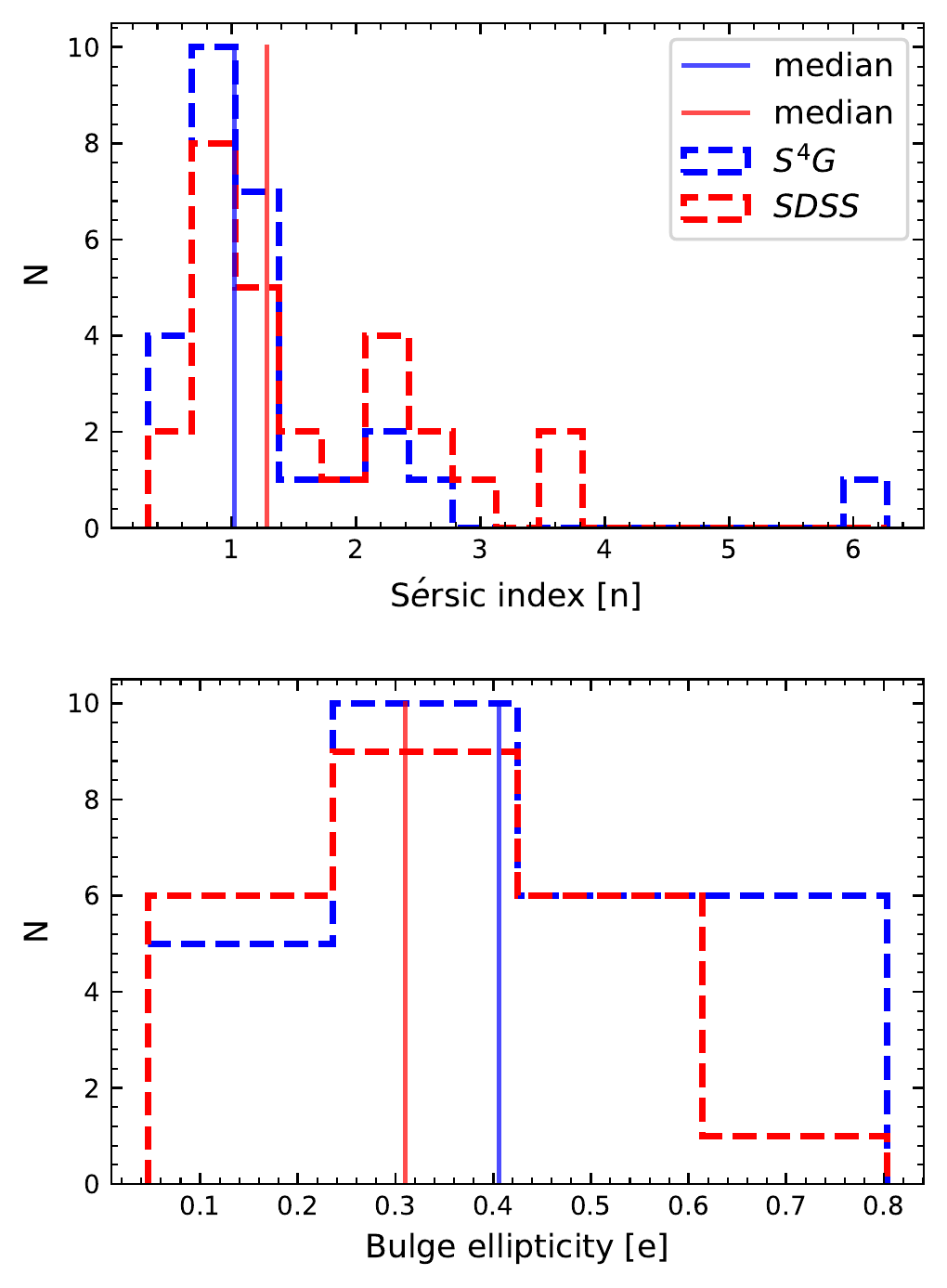}
    \caption{Comparison of SDSS survey with $S^4G$ survey. Top panel shows the S\'ersic index histogram and bottom panel shows the bulge ellipticity histogram of 27 galaxies, common in both survey, which are fitted by S\'ersic component. The red and blue filled histograms show the galaxies from SDSS and $S^4G$ survey respectively and vertical lines represent the medians of respective data. SDSS data over-estimates S\'ersic index and under-estimates bulge ellipticity.}
    \label{fig:compare_sdss_s4g}
\end{figure}

The SDSS data provides two-component fittings of the galaxies. But, in reality, a galaxy can have more or less than two components. Therefore, two-component fittings can provide biased parameters of the galaxies which do not have exactly two components \citep{Aguerri.etal.2005, Laurikainen.etal.2005, Gadotti.2009, Weinzirl.et.al.2009, Mendez-Abreu.etal.2014, Mendez-Abreu.etal.2017, Gao.etal.2019}. So, it is always good to go for the multi-component fittings of the galaxies. The robustness of the multi-component fittings over the two-component fittings has been discussed in \cite{Salo2015} for barred and unbarred galaxies. The two-component fittings of barred galaxies over estimate the mass, size and S\'ersic index of the bulges but there is no effect on unbarred galaxies. In this subsection, we will use $S^4G$ data to investigate the effect of multi-component fittings on our results. In the Fig.~\ref{fig:compare_sdss_s4g}, we have shown the histograms of S\'ersic index (top panel) and bulge ellipticity (bottom panel) for 27 galaxies, common in both survey, which are fitted by S\'ersic profile. The red and blue filled histograms show the galaxies from SDSS and $S^4G$ survey respectively and vertical lines represent the medians of respective data. We removed all the galaxies having bar and/or point source while making these histograms. The median of S\'ersic indices is 1.02 (standard deviation = 1.16) for $S^4G$ galaxies and 1.28 (standard deviation =0.95) for SDSS galaxies. On the other hand, the median of bulge ellipticity for $S^4G$ is 0.41 (standard deviation = 0.21) and for SDSS is 0.31 (standard deviation =0.19). We can clearly see the clues of over-estimation of S\'ersic index and under estimation of bulge ellipticity in SDSS data. In the estimation of structure parameters of galaxies, GALFIT is better than the GIM2D \citep{Haussler.etal.2007}. Hence, we can say that our result will not be affected by the multi-component fitting rather it will stand in the support of our results with much better confidence that the fraction of pseudo-bulges is increasing as Universe is getting older and older.

\begin{figure}
    \centering
    \includegraphics[width=\columnwidth]{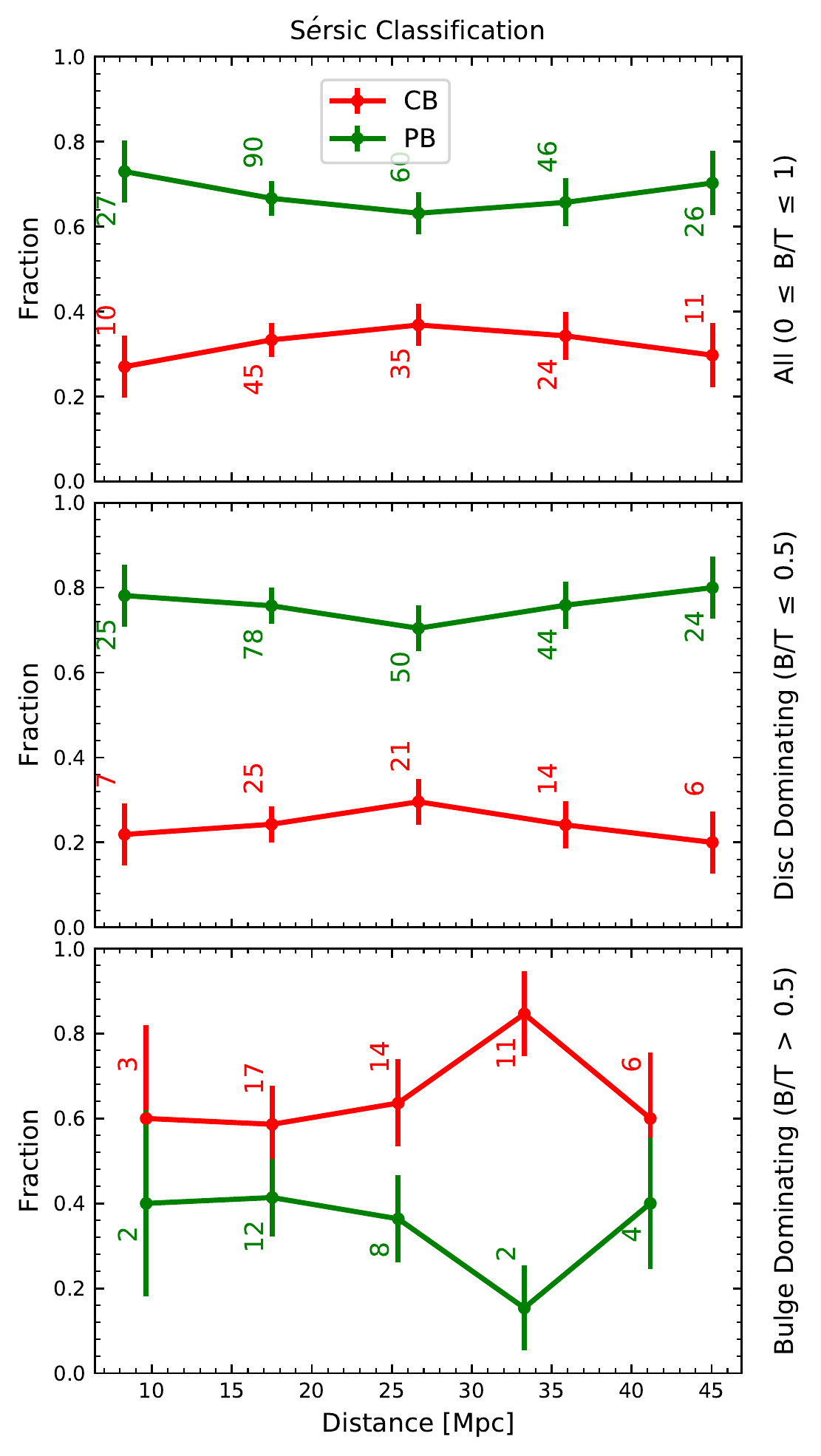}
    \caption{Fractional distribution of classical and pseudo-bulges with distance in the $S^4G$ survey. Top, middle, and bottom rows represent all, disc dominating, and bulge dominating galaxies respectively. Red and green curves represent classical bulges and pseudo-bulges respectively. The numbers of classical bulge and pseudo-bulge galaxies at any point are shown with their respective colors. Average fraction of the bulges here is equivalent to the local volume bulge fraction in SDSS data.}
    \label{fig:dist_bulge_s4g}
\end{figure}

We have also calculated the distribution of the bulges with distance for $S^4G$ survey after removing galaxies having bar and/or point source. Fig.~\ref{fig:dist_bulge_s4g} shows the distribution of classical and pseudo-bulges in the local Universe for $S^4G$ survey. Top, middle, and bottom rows of this figure represent all, disc dominating, and bulge dominating galaxies respectively. Red and green curves represent classical bulges and pseudo-bulges respectively. Here, we have shown only S\'ersic classification of the bulges because all three classifications give over all similar qualitative trends with small quantitative difference. From top two panels of the figure, one can easily notice that the pseudo-bulges dominates over classical bulges over whole range of distance. This domination is similar to the SDSS data in local volume. On comparison in very local volume, it seems that $S^4G$ data show larger fraction of pseudo-bulges than the SDSS data. This is because all the points in $S^4G$ data are equivalent to one point in SDSS data of the Fig.~\ref{fig:bulge_with_z}. If we take average of all the $S^4G$ points, it will give comparable fraction as that in SDSS data. For example, average fraction of pseudo-bulges in all $S^4G$ data is 0.66, which is very close to the fraction of pseudo-bulges in all SDSS data within statistical uncertainties.

In the bulge dominating galaxies (bottom panel of Fig.~\ref{fig:dist_bulge_s4g}), the fraction of pseudo-bulges is less than the fraction of classical bulges i.e classical bulges are dominating over pseudo-bulges. This is opposite to that in upper two panels but, it is equivalent to the fraction of bulges in bulge dominating galaxies of SDSS data. If we look at top and middle panels of  Fig.~\ref{fig:dist_bulge_s4g}, we can see that both of them are showing more or less similar fractions of bulges. It means that the over all fractional distribution of the bulges in $S^4G$ data is governed by the disc dominating galaxies not by the bulge dominating galaxies. We found that the data which is fitted with S\'ersic profile has more than $78\%$ disc dominating galaxies. Since the $S^4G$ survey is limited by the size and magnitude of the galaxies \citep{Sheth2010}. Therefore the fractional distribution of the bulges in the bottom panel is biased by the survey limitations. We are most probably seeing the merger dominated massive galaxies.

\section{Effect of Disc Inclination and Instrument's Resolution}
\label{sec:effect_of_inc_res}
\begin{figure}
    \centering
    \includegraphics[width=\columnwidth]{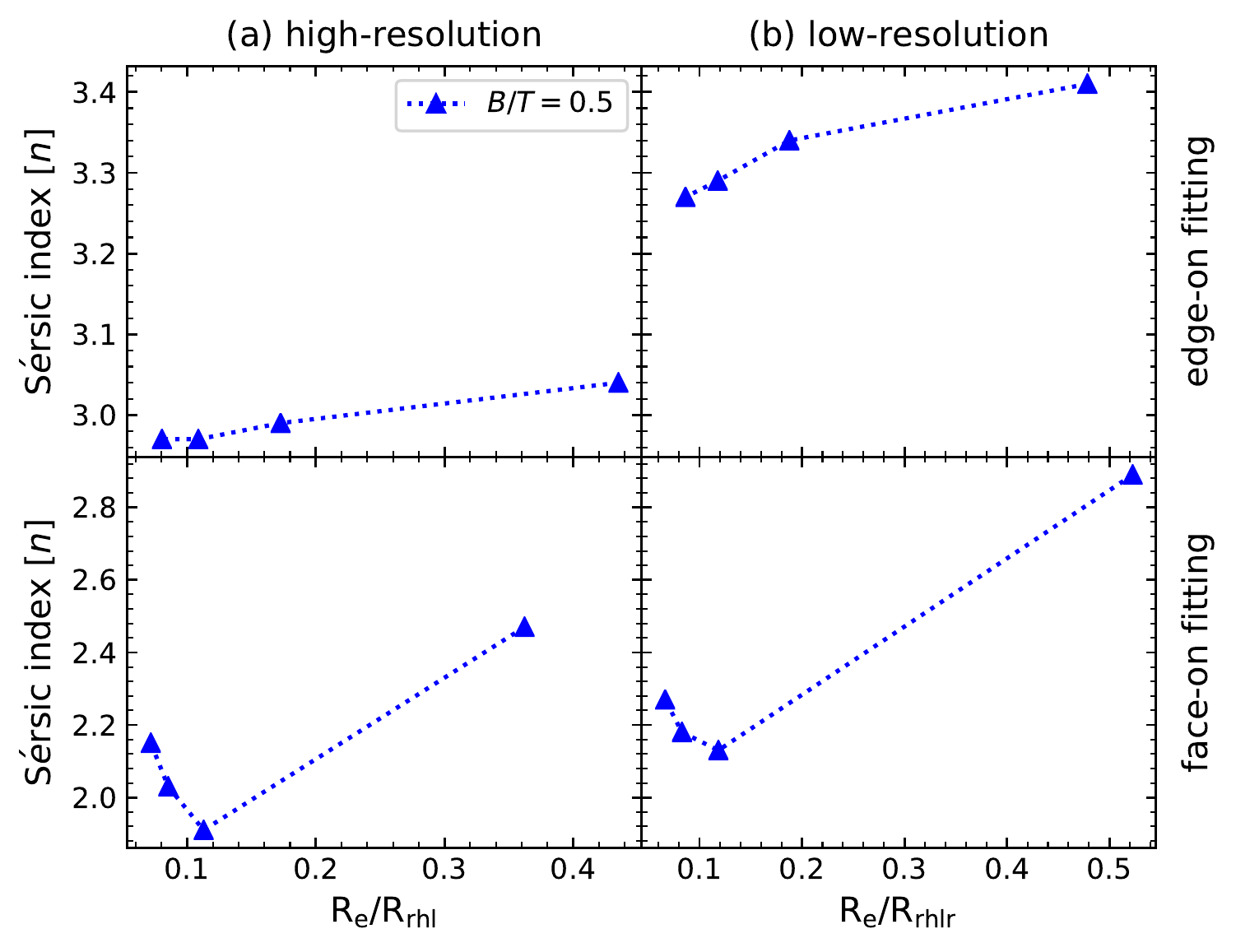}
    \caption{Effect of disc inclination and telescope's resolution on classical bulges of model galaxies. Left and right columns show high-resolution and low-resolution galaxies respectively. Top and bottom rows represent edge-on and face-on fittings of the model galaxies respectively. The blue upper triangle solid curves represent the models $B/T=0.5$ respectively and dots are just connecting points.}
    \label{fig:re_rhlr_ns_model}
\end{figure}

The state of art cosmological simulations and observational studies suggest that the Universe was dominated by mergers in the early epoch of galaxy formation \citep{White.Rees.1978, Springel.etal.2006, Sinha.etal.2012}. The violent relaxation of merged galaxies usually lead to the formation of classical bulges \citep{Kauffmann.et.al.1993, Hopkins.et.al.2009, Naab.et.al.2014}. Once formed, it is not an easy task to transform a classical bulge into a pseudo-bulge \citep{Kumar.etal.2021} unless there is a central instability (e.g. bar) which can make it dynamically cool \citep{kanak.etal.2012, Kanak.etal.2016} by transferring significant angular momentum \citep{Kataria.Das.2019}. But, instead of being dominated by classical bulges, the local Universe is dominated by the pseudo-bulges. Where have classical bulges gone in the evolution of galaxies? Is there any role of the two-component fitting? How do the disc surface density and inclination affect the two-component fitting? In the search of these questions as well as to test the reliability of S\'ersic index method, we have generated $N-$body disc galaxies with classical bulges and performed two-dimensional bulge-disc decomposition using GALFIT. 

Fig.~\ref{fig:re_rhlr_ns_model} shows the effect of disc inclination (rows) and spatial resolution (columns) on the S\'ersic index of classical bulges present in discs with a range of surface densities or in a range of R$_{\rm e}$/R$_{\rm hlr}$. In this figure, blue upper triangle solid curves represent the galaxy model with $B/T=0.5$. We have conducted a similar analysis for several values of $B/T$. However, we show only one for brevity, given our focus on the importance of disc surface density with respect to the central bulge. In this section, we show that only the S\'ersic index is not a reliable quantity to separate two classes of the bulges. We emphasize these points because it is one of the two parameters used in our new S\'ersic-Kormendy classification of bulges. At a given telescope's resolution, the S\'ersic index in edge-on fittings is always greater than the S\'ersic index in face-on fittings. The edge-on decomposition of the model galaxies provides the same S\'ersic index as reported by \cite{Dehnen1993} for pure Hernquist profile. The reasons for this difference between face-on and edge-on fitting are (1) Hernquist density profile which is not truncated at large radius in our models and (2) the real size of bulges can be traced only in edge-on projection of galaxies. Hence, the S\'ersic indices of the classical bulges depend on the disc inclination.

Each curve in Fig.~\ref{fig:re_rhlr_ns_model} represents the dependence of the S\'ersic index of a given classical bulge on the size (or surface density) of the hosting disc. The S\'ersic index of edge-on galaxies increases with increasing R$_{\rm e}$/R$_{\rm hlr}$ or we can say that the S\'ersic index of edge-on galaxies increases with decreasing size of hosting disc because all the bulges have same scale radius in our models. But the S\'ersic index of face-on galaxies first decreases and then increases with increasing R$_{\rm e}$/R$_{\rm hlr}$. In other words, one can think that the classical bulges easily pop-up in the low surface density discs. Similar results have been noticed in Fig.~\ref{fig:bulge_with_ReRd}, where we saw decreasing classical bulge fraction with increasing R$_{\rm e}$/R$_{\rm hlr}$ and then again increasing. The spatial resolution of the telescope also plays a crucial role in the determination of S\'ersic index. The low-resolution image usually takes large number of stars in each pixel which increases central density of the galaxy. This increased central density mainly contributes to the bulge in the term of increasing S\'ersic index. Therefore, the S\'ersic index in low-resolution fittings is always greater than the high-resolution fittings. These results have direct significance on the fraction of the bulges at high-redshift where the spatial resolution of the telescope decreases by a factor of distance. In Fig.~\ref{fig:bulge_with_z}, Some of the classical bulges at high-redshift could be the result of decreasing spatial resolution of the telescope.

\section{Effect of data selection criteria}
\label{sec:effect_of_data_selection}
\begin{figure*}
    \centering
    \includegraphics[width=\textwidth]{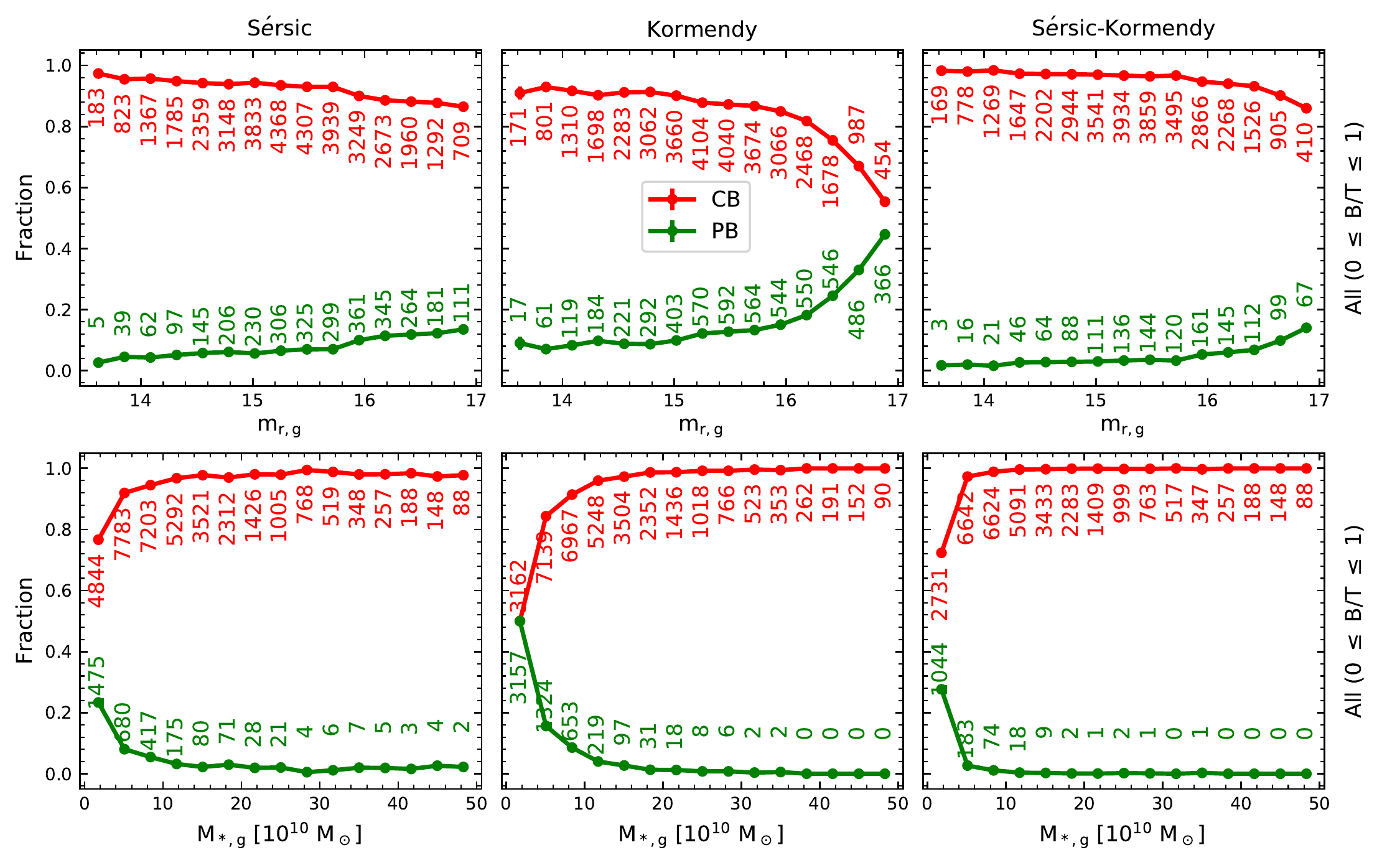}
    \caption{Fractional distribution of classical and pseudo-bulges with r-band apparent magnitude (m$_{\rm r,g}$) and stellar mass (M$_{*,g}$) of all the galaxies in the SDSS survey. Left, middle, and right columns show S\'ersic, Kormendy and S\'ersic-Kormendy classifications respectively. Red and green curves represent classical bulges and pseudo-bulges respectively. The numbers of classical bulge and pseudo-bulge galaxies at any point are shown with their respective colors. The fraction of pseudo-bulges increase in faint and low mass galaxies.}
    \label{fig:bulge_with_mag_mass}
\end{figure*}
The results, we have presented for SDSS data, are based on a sample of galaxies which are best represented by two components with least uncertainty. Imposing this strict constraint on the parent data reduces the galaxies from faint and low-mass end of the distribution, see Fig.~\ref{fig:sample_sdss}. Now, one can ask what will be the effect of this bias on our results? To understand the effect of reducing faint and low-mass galaxies from parent sample, we calculated the distribution of classical bulges and pseudo-bulges as a function of apparent magnitude and stellar mass of galaxies. It is shown in Fig.~\ref{fig:bulge_with_mag_mass}. The left, middle, and right columns of this figure show the distribution of bulges based on S\'ersic, Kormendy, and S\'ersic-Kormendy classifications respectively. Top row shows the fraction of bulges with r-band apparent magnitude of galaxies, whereas bottom row displays fractional distribution of bulges with stellar mass of the galaxies. The number of classical bulge and pseudo-bulge galaxies at any point of magnitude and stellar mass are also shown with respective colors.

From Fig.~\ref{fig:bulge_with_mag_mass}, it is clear that the fraction of the pseudo-bulges increases when we move towards the faint and low-mass end of galaxies in all the classification schemes. On the other hand, the fraction of classical bulges increases when we go from faint and low-mass end to bright and high-mass end of the galaxies. If we had not imposed our strict selection criteria, we would have seen increase in pseudo-bulge fraction as compare to that reported in Section~\ref{sec:bulge_with_z}. We have verified it for parent sample but, for the brevity, is not shown here. Also, the inclusion of faint and low-mass galaxies will not affect our results discussed in Section~\ref{sec:ellipticity_with_z} as those results are based on the mean of the bulge ellipticity of statistically significant number of galaxies. Finally, we would like to emphasize that including large number of galaxies will not affect our results though it will only reduce the statistical uncertainties of our results.

\section{Discussion}
\label{sec:discussion}
In previous section, we have presented our results on the evolution of the bulges since $z=0.1$ redshift. The effect of instrument's spatial resolution and galaxy inclination is shown using Milky Way type model galaxies. The correlation of the bulges with various photometric parameters and effect of multi-component fitting of the galaxy is also discussed. In this section, we interpret our results on the evolution of the bulges in observations.

The classification of bulges in classical bulge and pseudo-bulge categories has always been the topic of debate. Different people use different dividing criteria which suite their needs e.g $n=2$ S\'ersic index \citep{Fisher2006_proce, Fisher.Drory.2008}, Kormendy relation \citep{Gadotti.2009}, Color index \citep{Lackner.Gunn.2012}, velocity dispersion \citep{Fabricius.etal.2012}, central stellar mass density \citep{Cheung.etal.2012}, etc. We defined bulges into classical and pseudo categories using a combination of S\'ersic index and Kormendy relation which we call S\'ersic-Kormendy classification and compared properties of the bulges in three classification schemes. We found that the results based on S\'ersic classification and Kormendy classification deviate more from each other with increasing bulge to disc semi-major axis ratio. Results from combined S\'ersic-Kormendy classification lies in between those from two individual methods.

The fraction of classical bulges and pseudo-bulges varies with the evolution of the Universe. As we go from low-redshift to high-redshift, the fraction of pseudo-bulges decreases while the fraction of classical bulge increases. At redshift $z \approx 0.016$ for S\'erisc-Kormendy bulge classification, both types of bulges contribute equally. Pseudo-bulges are thought to be formed due to secular evolution of galaxies. The ubiquitous distribution of pseudo-bulge hosting disc galaxies in the local volume raise the question at hierarchical nature of the Universe \citep{Weinzirl.et.al.2009, Kormendy.et.al.2010, Fisher.Drory.2011}. Since the existence of zoom-in and hydrodynamic simulations, many studies have shown the importance of gas against the destruction (or heating) of the disc in merger events. These studies show that the pseudo-bulge can also form in mergers in contrary to the secular evolution \citep{Governato.etal.2009, Hopkins.etal.2009b, Moster.etal.2010, Guedes.etal.2013, Okamoto2013}. Other studies suggest to improve the initial condition of galaxy formation \citep{Peebles.2020}. A proper understanding of the physical processes and/or the adjustment to the current cosmological model is still needed for better representation of the observable Universe.

We found that a large fraction of the pseudo-bulges are rarer in density as compare to the classical bulges relative to their hosting discs. \cite{Gadotti.2009} also found that the pseudo bulges less are concentrated than the classical bulges at given bulge to disc mass ratio. High stellar density discs usually go through instabilities e.g. bars. \cite{Lutticke.etal.2000} studied a sample of edge-on galaxies from RC3 catalogue (Third Reference Catalogue of Bright Galaxies) and found that $45\%$ of all bulges are boxy/peanut in shape and explained by the presence of bar. Rare density of pseudo-bulges is in agreement with previous studies where \citep{Erwin.Debattista.2017} claims that the semi-major axes of boxy/peanut shape bulges range from one-quarter to three-quarters of the full bar size. If the classical bulges are small in mass, these instabilities can hide them in (or sometime erode completely) and can result in composite bulges as seen in some observations \citep{Erwin.et.al.2015, BlanaDiaz.etal.2018, Erwin.etal.2021}. The increasing fraction of pseudo-bulges with decreasing redshift could be the result of these hidden low mass classical bulges \citep{Kanak.2015}. Further, our result shows the bulge dominated systems also have higher fraction of pseudo bulges at low-redshift which is opposite to earlier claims \cite{Gadotti.2009}. Though our results supports the composite bulge scenario. A detailed multi-component decomposition of galaxies can reveal what fraction of the pseudo-bulges are composite bulges. Further these techniques can certainly help in improving our understanding of the galaxy evolution. 

The two-dimensional photometric decomposition of the galaxies gives not only the quantitative information of the bulge and disc components but, it also provides the tool to understand the evolution of bulges. From the sub-sample of near face-on galaxies in SDSS data, we found that massive bulges are more round than the low mass bulges at all redshift. In local volume, classical bulges show mean ellipticity in the range 0.2 to 0.3, whereas pseudo-bulges show in the range 0.3 to 0.5 for S\'ersic-Kormendy bulge classification which is consistent with previous studies of spiral galaxies \citep{Fathi.Peletier.2003, Mendez-Abreu.etal.2008}. Also, the mean ellipticity of both types of bulges grows with the increasing redshift. It indicates a clear morphological evolution of the bulges since $z=0.1$. The decreasing mean ellipticity and increasing pseudo-bulge fraction point towards the secular evolution in the later time of Universe. Recent cosmological simulations have also shown that the rate of violent interactions (mergers) between galaxies decreases with the evolution of the universe and slowly, flyby interactions dominate the evolution \citep{Sinha.etal.2012, An.etal.2019}.

The spatial resolution of the observing instrument also play a crucial role in the identification of the bulge type Fig.\ref{fig:re_rhlr_ns_model}. At low spatial resolution, a pseudo-bulge can be mistakenly classified as classical bulge in photometric decomposition of the bulge using S\'ersic profile. Since the spatial resolution of the instrument decreases when we see a high-redshift object. Hence, the fraction of the pseudo-bulge at high-redshift could be higher than what we have calculated from the sample of photometric bulge-disc decomposition. A further investigation is needed to tightly constrain the fraction of the cold and hot fraction of stellar matter in the Universe at different redshift. Calculation of the exact fraction of ordered and random stellar orbits in disc galaxies at different redshift will help us in better understanding of the initial cosmological conditions and/or the underlying baryonic physics.

Our study is based on two-component photometric decomposition. The effect of multi-component decomposition of the galaxies is very crucial to understand the real statistics of the stellar matter distribution in various components of the galaxy. Two-component fitting usually gives higher S\'ersic index and lower ellipticity in case of the barred galaxies (Figure \ref{fig:compare_sdss_s4g}). This can lead to small fraction of the pseudo-bulges and elongated bulges. Sometimes, photometric and kinematic decomposition of cold and hot components show contrary classification of the bulges \citep{Gadotti.2009}. For example, the excess light in the center of galaxy due to nuclear activity gives higher S\'ersic index in photometric decomposition. The detailed relation in photometric and kinematic decomposition is necessary for better photometric classification criterion.

\section{Summary}
\label{sec:summary}
To understand the evolution of the bulges with the evolution of the Universe, we have used the archival data of two-dimensional bulge-disc decomposition of galaxies from SDSS DR7 \citep{Simard2011}. This data is constraint to nearly 40000 galaxies those can be well represented by two components. The classical bulges and pseudo-bulges are separated using $n=2$ S\'ersic index, Kormendy relation, and a combination of two. To explore the effect of multi-component fitting, we have also used the archival data of two-dimensional multi-component decomposition of local volume galaxies from $S^4G$ survey \citep{Salo2015}. Further, we have simulated Milky Way type model galaxies to investigate the effect of spatial resolution and disc inclination. The main findings of our analysis are as follows:-

(i) The fraction of pseudo-bulges dominates over classical bulges in the local Universe. As we go towards the high-redshift, the fraction of pseudo-bulges decreases smoothly. In the history of the Universe, there came a point (z $\approx$ 0.016 for S\'ersic-Kormendy classification) when the fraction of classical bulges and pseudo-bulges was 50-50$\%$. Universe is dominated by classical bulged galaxies before this point during its evolution while it is dominated by pseudo-bulges as soon it crosses this point.

(ii) In the local Universe, disc dominating galaxies show more pseudo-bulges as compare to bulge dominating galaxies. The fractional distributions of pseudo-bulges in Kormendy, and S\'ersic-Kormendy classifications are quite similar. But, S\'ersic classification shows lower pseudo-bulges than other two classification at low-redshift.

(iii) The fraction of the pseudo-bulges increases with increasing bulge to disc semi-major half-light ratio until $R_{\rm e}/R_{\rm hlr} \approx 0.6$. Compact and shorter bulges, as compared to their hosting discs, are usually classical in nature however, pseudo-bulges are diffuse and longer. In other words, concentrated discs harbour pseudo-bulges while rarer discs host classical bulges.

(iv) At large bulge to disc semi-major axis ratio, Kormendy classification shows more pseudo-bulges than other two classifications. This difference is more pronounced in disc dominating galaxies. Bulge dominating galaxies show very similar fraction in all classifications. For better division between classical and pseudo-bulges, we recommend to use a combination of S\'ersic index and Kormendy relation based bulge classification.

(v) The mean ellipticity of pseudo-bulges is greater than the mean ellipticity of classical bulges in whole redshift range and it decreases with decreasing redshift indicating that the bulges are getting more axisymmetric with the evolution of galaxies. High-mass bulges are progressing towards the axisymmetry at more steep rate than the low-mass bulges.

(vi) In local Universe, nearly 75$\%$ of the visible matter is dominated by ordered rotational motion in disc galaxies. The existence of the rotation dominated Universe challenges the hierarchical nature of the Universe which suggests that most of the Universe should be dispersion dominated.

(vii) The evolution of the bulge does not have significantly strong correlation with the evolution of the photometric properties of galaxy e.g. bulge to total light ratio, disc inclination, bulge ellipticity, absolute magnitude of galaxy etc. Only absolute magnitude of galaxies and ellipticity of bulges show a negative correlation of $\approx 0.3$ on the scale of unity in the low-redshift Universe that also drops in very local Universe.

(viii) Our results are consistent with the local volume survey $S^4G$. The multi-component fitting of the galaxies does not fade our conclusions. It stands with strong support of our results discussed in this article. 

The tight constraint on the fraction of the classical bulges and pseudo-bulges with the redshift can be verified with next generation telescopes e.g. JWST, TMT, SKA etc. These telescope will provide unprecedented high-resolution view of the universe. Using the data obtained with these next generation telescope, we can better understand the underlying mechanism which played major role in the stability of discs during the growth and evolution of the structures.

\section{Acknowledgements}
We thank anonymous reviewer for useful comments and suggestions that enhanced the quality of our article. We thank Sudhanshu Barway and Vivek M for the insightful discussions during this work. AK thanks Mousumi Das for her support. This research has made use of the VizieR catalogue access tool, CDS, Strasbourg, France (DOI: 10.26093/cds/vizier). The original description of the VizieR service was published in A$\&$AS 143, 23. Various Python based packages e.g. NumPy \citep{numpy2020}, Matplotlib \citep{matplotlib2007}, and Astropy \citep{astropy2018}, and Scipy \citep{Scipy2020} have been used in this study.

\section{Data Availability}
The data underlying the present article are available in VizieR Catalogue, at \href{https://doi.org//10.26093/cds/vizier.21960011}{https://doi.org//10.26093/cds/vizier.21960011} \citep{Simard2011} and \href{https://doi.org//10.26093/cds/vizier.22190004}{https://doi.org//10.26093/cds/vizier.22190004} \citep{Salo2015}. Galaxy Zoo data on classification of galaxies is available at \href{https://data.galaxyzoo.org/}{https://data.galaxyzoo.org/} \citep{Willett.etal.2013}.

\bibliographystyle{mnras}
\bibliography{bib.bib}

\appendix
\section{Comparison of Bulges and Discs}
\label{app:z_vs_Re_Rd}
\begin{figure*}
    \centering
    \includegraphics[width=\textwidth]{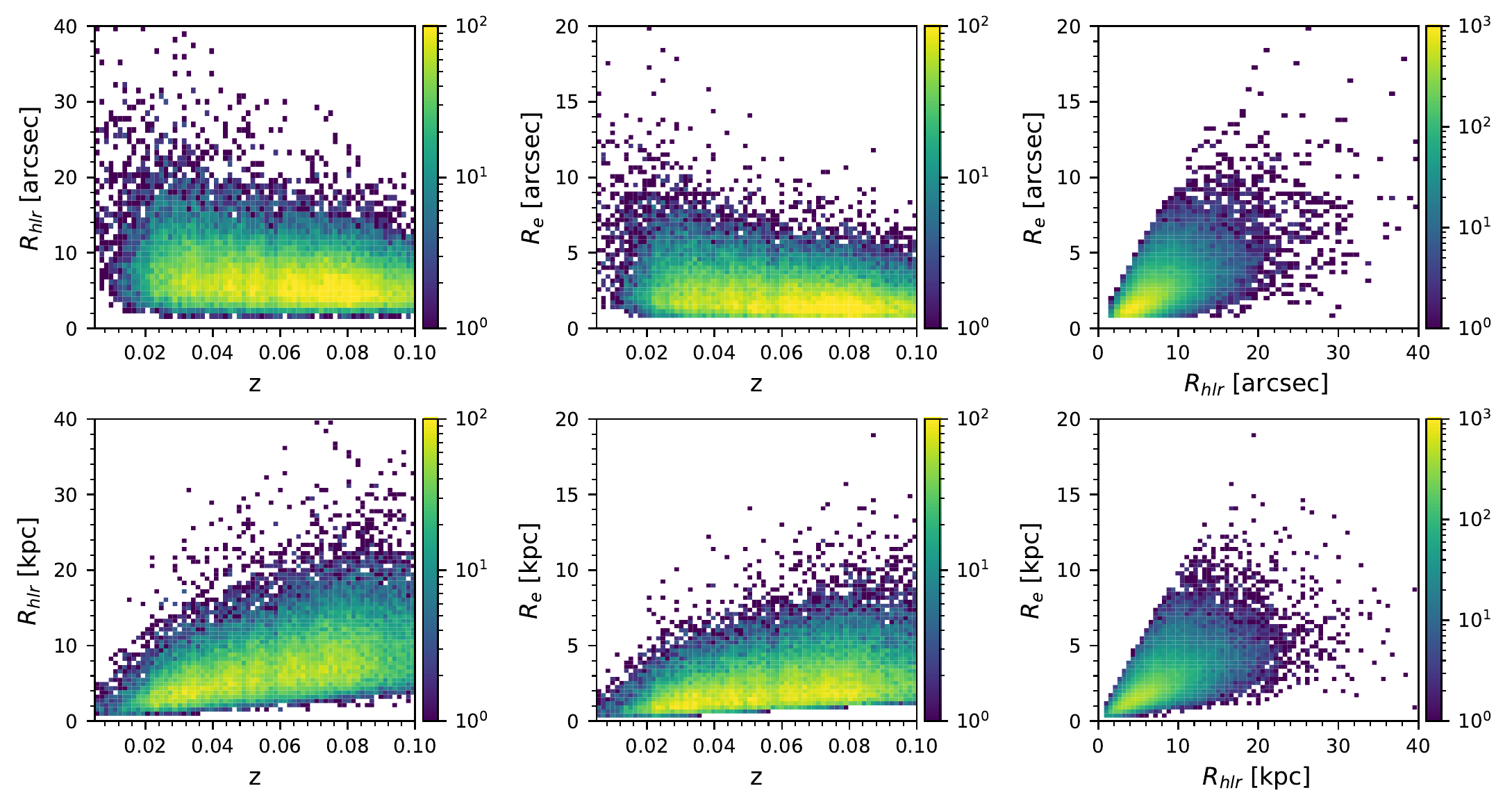}
    \caption{Sizes of the bulges and discs in our sample. Left and middle columns demonstrate absolute sizes of discs and bulges respectively as a function of redshift, whereas right column shows their comparison. Top and bottom rows display apparent sizes and physical sizes respectively.}
    \label{fig:z_vs_Re_Rd}
\end{figure*}

In Fig.\ref{fig:z_vs_Re_Rd}, we have shown the apparent and physical sizes of the bulges and discs as a function of redshift. Top and bottom rows show apparent sizes and physical sizes respectively. Left and middle columns display absolute sizes of discs and bulges respectively, whereas right column shows their comparison. There are some galaxies that fall out of the shown range of the axes. However, for better visualization, range of the axes are chosen arbitrarily. As expected, apparent sizes of the bulges and discs decrease with increasing redshift, while physical sizes increase with increasing redshift. Comparison of bulges and discs show notable positive correlation. On average, larger discs host larger bulges and smaller discs hot smaller bulges. This correlation shows increasing dispersion with increasing disc size.

\section{Bulge to Disc Central Brightness}

\label{app:bulge_disc_bright}
\begin{figure}
    \centering
    \includegraphics[width=\columnwidth]{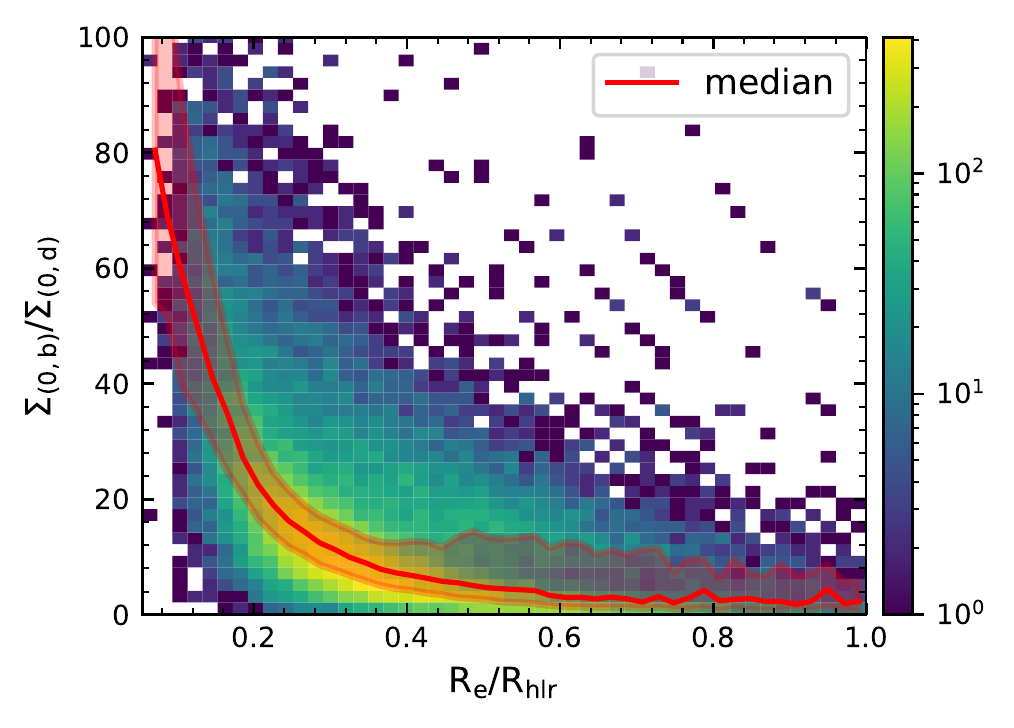}
    \caption{Distribution of the bulge to disc central brightness ratio for the sample selected from SDSS survey with the bulge to disc semi-major half-light radius. The red curve shows the median of the distribution and shaded region represents $\pm 25\%$ deviation from median.}
    \label{fig:bulge_disc_bright}
\end{figure}

In Fig.~\ref{fig:bulge_disc_bright}, we have shown the ratio of bulge central brightness ($\Sigma_{(0,b)}$) to the disc central brightness ($\Sigma_{(0,d)}$) as a function of the bulge to disc semi-major half-light radius ratio. This ratio is calculated using the following expression,
\begin{equation}
    \frac{\Sigma_{(0,b)}}{\Sigma_{(0,d)}}=\frac{B}{D} \times \frac{R_{hlr}^2}{R_{e}^2}
    \label{eqn:bulge_disc_bright}
\end{equation}
where, $\frac{B}{D}$ is the photometric bulge to disc light ratio obtained from two-component decomposition of the galaxies. $R_{\rm e}$ and $R_{\rm hlr}$ are the semi-major half-light radii of the bulge and disc components respectively. This figure is clearly showing that the bulge concentration is decreasing with increasing bulge to disc semi-major half-light radius ratio. It is evident that the $R_{\rm e}/R_{\rm hlr}$ can be used as an indicator to measure the bulge concentration relative to the hosting disc.

\end{document}